\documentclass[pageno]{jpaper}
\pdfoutput=1 

\usepackage[normalem]{ulem}
\usepackage{tikz}
\usepackage{amsmath}

\usepackage{filecontents}
\usepackage{algorithmic}
\usepackage{graphicx}
\usepackage{textcomp}
\usepackage{xcolor}
\usepackage{enumitem}
\usepackage{xspace}
\usepackage{multirow}
\usepackage{comment}
\usepackage{threeparttable}
\usepackage{xstring}
\usepackage{booktabs}
\usepackage{authblk}

\begin{document}

\newcommand{\name}{MTM\xspace}
\definecolor{jie2}{RGB}{0, 0, 0}
\definecolor{jie}{RGB}{0,0,0}
\definecolor{dong}{RGB}{0,0,0}
\definecolor{check}{RGB}{0,0,0}
\definecolor{revision}{RGB}{0,0,0}
\title{Rethinking Memory Profiling and Migration for Multi-Tiered Large Memory Systems}

\author[1]{Jie Ren}
\author[1]{Dong Xu}
\author[3]{Ivy Peng}
\author[2]{Junhee Ryu}
\author[2]{Kwangsik Shin}
\author[2]{Daewoo Kim}
\author[1]{Dong Li}
\affil[1]{University of California, Merced}
\affil[2]{Sk Hynix}
\affil[3]{KTH Royal Institute of Technology}

\thispagestyle{empty}
\date{}
\maketitle
\begin{abstract}

Multi-terabyte large memory systems are often characterized by more than two memory tiers with different latency and bandwidth.
Multi-tiered large memory systems call for rethinking of memory profiling and migration because of the unique problems unseen in the traditional memory systems with smaller capacity and fewer tiers. We develop \name, an application-transparent page management system based on three principles: (1) connecting the control of profiling overhead with the profiling mechanism for high-quality profiling; (2) building a universal page migration policy on the complex multi-tiered memory for high performance; and (3) introducing huge page awareness. We evaluate \name using common big-data applications with realistic working sets (hundreds of GB to 1 TB). \name outperforms seven state-of-the-art solutions by up to 42\% (17\% on average).

\end{abstract}

\section{Introduction}
\label{sec:intro}

The memory hierarchy is adding more tiers and becoming more heterogeneous to cope with performance and capacity demands from applications. Multi-tier memory systems that started from multi-socket non-uniform memory access (NUMA) architecture is now a de-facto solution for building scalable and cost-effective memory systems. For instance, the Amazon EC2 High Memory Instance has three DRAM-based memory tiers built upon eight NUMA nodes~\cite{amazon_high_mem_inst}. 
Recently, the commercial availability of new memory technologies, such as high-bandwidth memory (HBM) and \textcolor{dong}{compute express link (CXL)~\cite{cxl},} 
is adding a new dimension to memory systems. As a result, a multi-tier memory system 
can easily exceed two memory tiers. Top tiers feature lower memory latency or higher bandwidth but smaller capacity, while bottom tiers feature higher capacity but lower bandwidth and longer latency. When high-density memory is in use, e.g., Intel's Optane DC PM~\cite{Optane:blogreview}, a multi-tier large memory system could enable high-performance, terabyte-scale graph analysis~\cite{vldb20_sage,optane_utexas19,peng2018graphphi}, in-memory database services~\cite{Andrei:2017:SHA:3137765.3137780,vldb21:ai_pm,ucsd_otpane:fast2020}, and scientific simulations~\cite{pm-octree:sc17,ics21:memoization} on a single machine in a cost-effective way.



Most of the page management systems for multi-tier heterogeneous memory (HM)~\cite{Agarwal:2017:TAP:3037697.3037706, intel_mem_optimizer, tiered-autonuma, atc21_autotiering,autonuma_balance, meta_tpp,sosp21_hemem} consist of three components -- a profiling mechanism, a migration policy, and a migration mechanism. A profiling mechanism is critical for identifying performance-critical data in applications and is often realized through tracking page accesses. A migration policy chooses candidate pages to be moved to top tiers. Finally, the effectiveness of a page management solution directly depends on whether its migration mechanism can move pages across tiers at low overhead. The emergence of multi-tiered large memory systems calls for rethinking of memory profiling and migration to address unique problems unseen in traditional single- or two-tier systems with smaller capacity.

\textbf{Problems.} 
The large memory capacity brings challenges to memory profiling. Linux and existing memory profiling mechanisms~\cite{autonuma}  manipulate specific bits in page table entries (PTEs) to track memory accesses at a per-page granularity. This profiling method has the benefit of application-transparency, but is not scalable on a large memory system. Our evaluation shows that tracking millions of pages could take several seconds -- too slow to respond to time-changing access patterns, and causes 20\% slowdown in TPC-C against VoltDB~\cite{voltdb}. The most recent solution DAMON~\cite{hpdc22:daos, middleware19_profiling,  damon} dynamically forms memory regions out of the virtual memory space mostly based on spatial locality, and profiles a single page per region. The total number of regions is constrained, such that the profiling overhead is controlled. DAMON has been adopted by Linux~\cite{damon}, and solves the profiling overhead problem faced by the large memory system, but its profiling quality can be out of control (shown in Section~\ref{sec:motivation}): DAMON can miss more than 50\% of frequently-accessed pages and is slow to respond to the variance of memory access patterns.

The limitation in profiling quality comes from (1) the rigid control over profiling overhead and (2) the ad-hoc formation of memory regions. Memory profiling relies on PTE scans. Given a large memory system with a certain constraint on profiling overhead, the PTE scan for memory profiling can only happen certain times. Deciding the distribution of those PTE scans in the memory regions is critical for effective profiling. Strictly enforcing one PTE scan per region (to profile one page), as in DAMON, breaks the functionality of the profiling mechanism and compromises profiling quality. Furthermore, the ad-hoc formation of memory regions 
\textcolor{dong}{(such as sometimes splitting each memory region to two)} 
takes a long time to find pages with the similar memory access patterns to form memory regions for profiling, delaying page migration.

In addition, rich memory heterogeneity brings challenges to page migration. Existing solutions, such as \textcolor{jie}{tiered-AutoNUMA}~\cite{tiered-autonuma} 
for multi-tiered memory are built upon an abstraction extended from the traditional NUMA systems, where page migration occurs between two neighboring tiers with the awareness of no more than two NUMA distances. However, such an abstraction limits multi-tiered memory systems, because migrating pages from the lowest to the top tier, at tier-by-tier steps, has to make multiple migration decisions to reach the destination tier, which takes multiple seconds and fails to timely migrate pages for high performance.

Furthermore, Linux and existing solutions do not consider the implications of huge pages on memory profiling and migrations. Using huge pages is common in the large memory systems to reduce TLB misses and avoid long traverse of page tables. The transparent huge page mechanism (THP) in Linux mixes huge pages and 4KB pages, which brings complexity to form memory regions for profiling. THP also calls for a support of effective migration of huge pages. 


\textbf{Solutions.} We argue that the following principles must be upheld to address the above problems. 

\begin{itemize}[leftmargin=*,noitemsep,topsep=0pt]
\item Connecting the control of profiling overhead with the profiling mechanism to enable high-quality profiling; 

\item Building a universal page migration policy on the complex multi-tiered memory hierarchy for high performance;

\item Introducing huge page awareness. 
\end{itemize}

In this paper, we contribute a page management system called \name (standing for \textit{M}ulti-\textit{T}iered memory \textit{M}anagement) that realizes the above principles on large multi-tier memory. 

\name decouples the control of profiling overhead from the number of memory regions, but connects it directly with the number of PTE scans (the profiling mechanism). Hence, profiling quality and overhead can be distributed proportionally according to the variation of both spatial and temporal locality. More PTE scans or page profiling can be enforced for a memory region where there is large variation hence demanding more fine-grained profiling. Also, the splitting of memory regions based on the variation is able to be guided rather than randomly \textcolor{dong}{happened} as in DAMON.

To avoid the slow formation of memory regions, \name uses performance counters to guide PTE scans in the largest memory tier. Only memory regions identified by performance counters due to memory events are subject to be profiled with higher accuracy and faster promotion to faster memory tiers. The performance counter-guided profiling is event-driven, providing promptness to catch changes in memory access patterns.


\name breaks the barrier that blocks the construction of a universal page migration policy across tiers. This barrier comes from the limited memory profiling functionality (either at the slowest NUMA node~\cite{meta_tpp} or random selection of hundreds of MB on NUMA nodes ~\cite{autonuma,mem_optimizer_intel, memory_tiering, atc21:autotiering, meta_tpp}). \name uses the overhead-controlled, high-quality profiling to establish a \textit{global} view of all memory regions in all tiers and consider all NUMA distances to decide page migration. In particular, \name enables a ``fast promotion and slow demotion'' policy for high performance.  Hot (frequently-accessed) pages identified in all lower tiers are ranked and directly promoted to the top tier, minimizing data movement through tiers. When a page is migrated out of the top tier to accommodate hot pages, the page is moved to the next lower tier with available space. This policy needs no apriori knowledge of the number of memory tiers in a system and makes the best usage of fast tiers. 

\name also features a fast migration mechanism, which is critical for realizing the migration policy in practice. This mechanism dynamically chooses from an asynchronous page copy-based scheme and a synchronous page migration scheme, based on the read/write pattern of the migrated page, to minimize migration time.

Finally, page migration and profiling in \name fully supports huge pages and THP, embodied as page alignment during splitting and merging of memory regions and effective support of huge page migration.


\textbf{Evaluation.} We rigorously evaluated \name against seven state-of-the-art solutions, including 
two state-of-the-art solutions (AutoTiering~\cite{atc21_autotiering} and HeMem~\cite{sosp21_hemem}), an existing solution in Linux \textcolor{check}({tiered-AutoNUMA}~\cite{autonuma}), a hardware-based solution (Optane Memory Mode), and first-touch NUMA. \name is also compared against two kernel-based page migration solutions (the ones in Linux and Nimble~\cite{Yan:2019:NPM:3297858.3304024}). \textcolor{revision}{\name outperforms Memory Mode, first-touch NUMA, tiered-AutoNUMA, AutoTiering, AutoNUMA and HeMem by 20\%, 22\%, 24\%, 25\% and 24\%.} \name outperforms the Linux and Nimble migration approaches by 40\% and 36\% for read-intensive workloads, and performs similarly for write-intensive workloads.

\section{Background and Related Work}
\label{sec:bg}

\begin{figure}[t!]
    \centering
        \includegraphics[width=1\columnwidth]{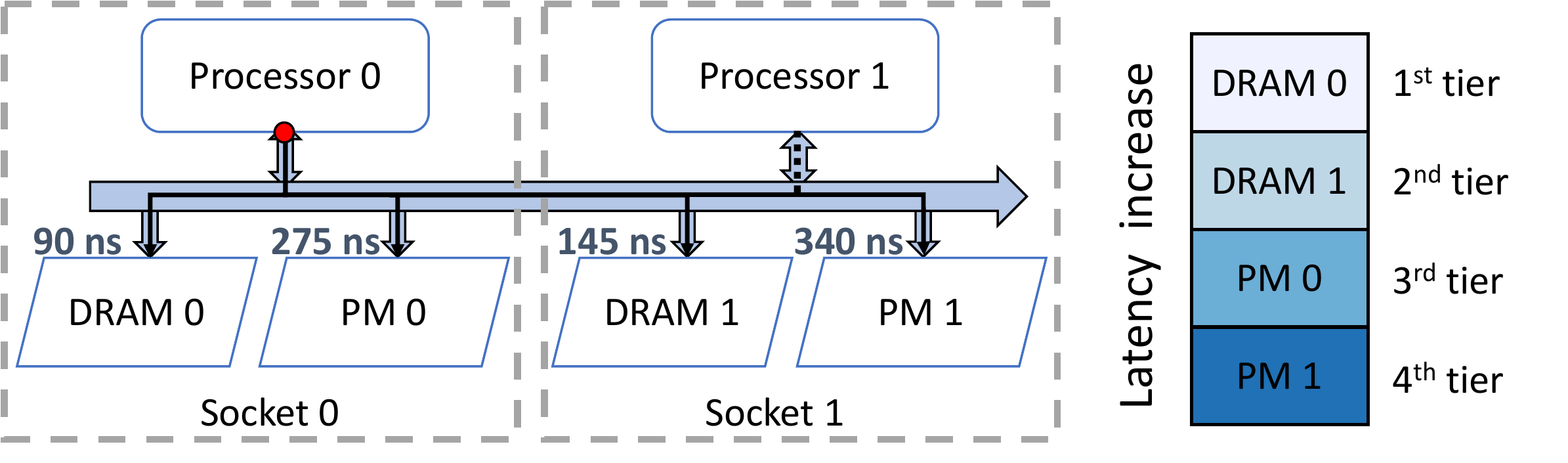}
        \vspace{-10pt}
    \caption{An example of multi-tiered memory system.}
    \label{fig:multi-tiered-mem}
\end{figure}

We introduce memory organization in multi-tiered large-memory systems. \textcolor{dong}{In such a system, each processor has its local memory as a fast memory tier, and has memory expansion or other processor's local memory as a tier of slow memory. The memory expansion may be based on CXL interconnect and appear as CPU-less memory nodes. Figure~\ref{fig:multi-tiered-mem} shows an example of such a system which is an Intel Optane-based system with two sockets and four memory components (i.e., DRAM 0-1 and PM 0-1). Two PM components appear as CPU-less memory nodes in Linux. Each process has its local DRAM as a fast memory tier, and has two PM components and a remote DRAM as three tiers of slow memory. Another example of such a system can be found from Microsoft Azure~\cite{ms_cxl_2022}.}





\subsection{Large Memory Systems}
Given emerging large memory systems, there is a pressing need of studying effectiveness and scalability of system software and hardware to support them. We review related works. 


\textbf{Software support for page management.} Recent work manages large memory systems based on an existing NUMA balancing~\cite{autonuma_balance} solution in Linux. \textcolor{dong}{Tiered-AutoNUMA~\cite{tiered-autonuma} periodically profiles 256MB memory pages. Tiered-AutoNUMA balances memory access between CPU-attached memory nodes, and then balance memory accesses between CPU-attached memory node and CPU-less memory node based on page hotness. As a result, a hot page takes a long time to migrate to the fastest memory for high performance.}
AutoTiering~\cite{atc21_autotiering} is a state-of-the-art solution. \textcolor{dong}{It uses the same profiling method 
as Tiered-AutoNUMA,} and introduces flexible page migration across memory tiers. However, it does not have a systematic migration strategy guided by page hotness. HeMem~\cite{sosp21_hemem} is a state-of-the-art solution for two-tiered PM-based HM. HeMem only uses performance counters to identify hot pages and fails to explore more than two tiers.

\textbf{Hardware-managed memory caching.} 
Some large memory systems use fast memory as a hardware-managed cache to slow memory. For example, in Intel's Optane PM, DRAM can work as a hardware-managed cache to PM in the \textit{Memory Mode}. However, this solution results in data duplication, wasting fast memory capacity. It also causes serious write amplification when there are memory cache misses~\cite{nvmw21:dram-cache}.




\subsection{Two-Tiered Heterogeneous Memory}
HM combines the best properties of memory technologies optimized for performance, capacity, and cost, but complicate memory management. There are application-transparent solutions that measure data reuse and migrate data for performance~\cite{Agarwal:2017:TAP:3037697.3037706, kleio:hpdc19,ipdps21_cori, Hirofuchi:2016:RHV:2987550.2987570, Kannan:2017:HOD:3079856.3080245, asplos21:kloc, asplos19_softwarefarmem,tpds19_tieredmem,sosp21_hemem,Yan:2019:NPM:3297858.3304024}. However, they can cause uncontrolled profiling overhead or low profiling quality, and are not designed for more than two memory tiers. 

There are application-specific solutions that leverage application domain knowledge to reduce profiling overhead, prefetch pages from slow memory to fast memory, and avoid slow-memory accesses. Those solutions include big data analysis frameworks (e.g., Spark~\cite{pldi19:panthera}), machine learning applications~\cite{AutoTM_asplos20, hpca21_sentinel,neurips20:hm-ann}, scientific computing~\cite{pm-octree:sc17,peng2018siena,ics21:memoization}, and graph analysis~\cite{vldb20_sage,optane_utexas19,peng2018graphphi}. These solutions show better performance than the application-transparent, system-level solutions, but require extensive domain knowledge and application modifications. \name is an application-transparent solution.

\subsection{Huge Page Support in Linux}
Linux supports fixed-sized huge pages, e.g., 2MB and 1GB, on x86 architectures. While it is possible to manage huge pages explicitly within applications using specific APIs, Transparent Huge Pages (THP) is the most common approach to use huge pages, because it requires no application modifications. When THP is enabled, the kernel automatically uses huge pages to satisfy large memory allocation with page alignment. Also, a kernel daemon  (\texttt{khugepaged}) runs periodically to detect if contiguous 4KB pages can be promoted into a huge page or vice versa, mostly based on access recency and frequency. THP is swappable by splitting a huge page into basic 4KB pages before swapping out. Thus, when existing works reuse the Linux default page swapping routines to migrate pages to a slower memory tier~\cite{atc21:autotiering, meta_tpp}, the huge page cannot arrive at the slower memory tier unless it is split into 4KB pages because of access recency and frequency (not because of migration).

\section{Motivation}
\label{sec:motivation}
High accuracy, low overhead profiling mechanism is key for managing multi-tiered large memory. We study the profiling methods in state-of-the-art works (Thermostat~\cite{Agarwal:2017:TAP:3037697.3037706}, AutoTiering~\cite{atc21_autotiering}, Linux's DAMON~\cite{ middleware19_profiling, damon}) and summarize their overhead and accuracy tradeoff as follows. AutoTiering randomly chooses 256 MB pages for profiling to detect hot pages. Both Thermostats and DAMON maintain a list of memory regions and randomly choose one page per region for profiling. Thermostats keeps all memory regions in a fixed size while DAMON dynamically splits and merges memory regions to improve profiling quality -- they control profiling overhead by changing the number of memory regions.

\begin{figure}[t!]
    \centering
        \includegraphics[width=\columnwidth]{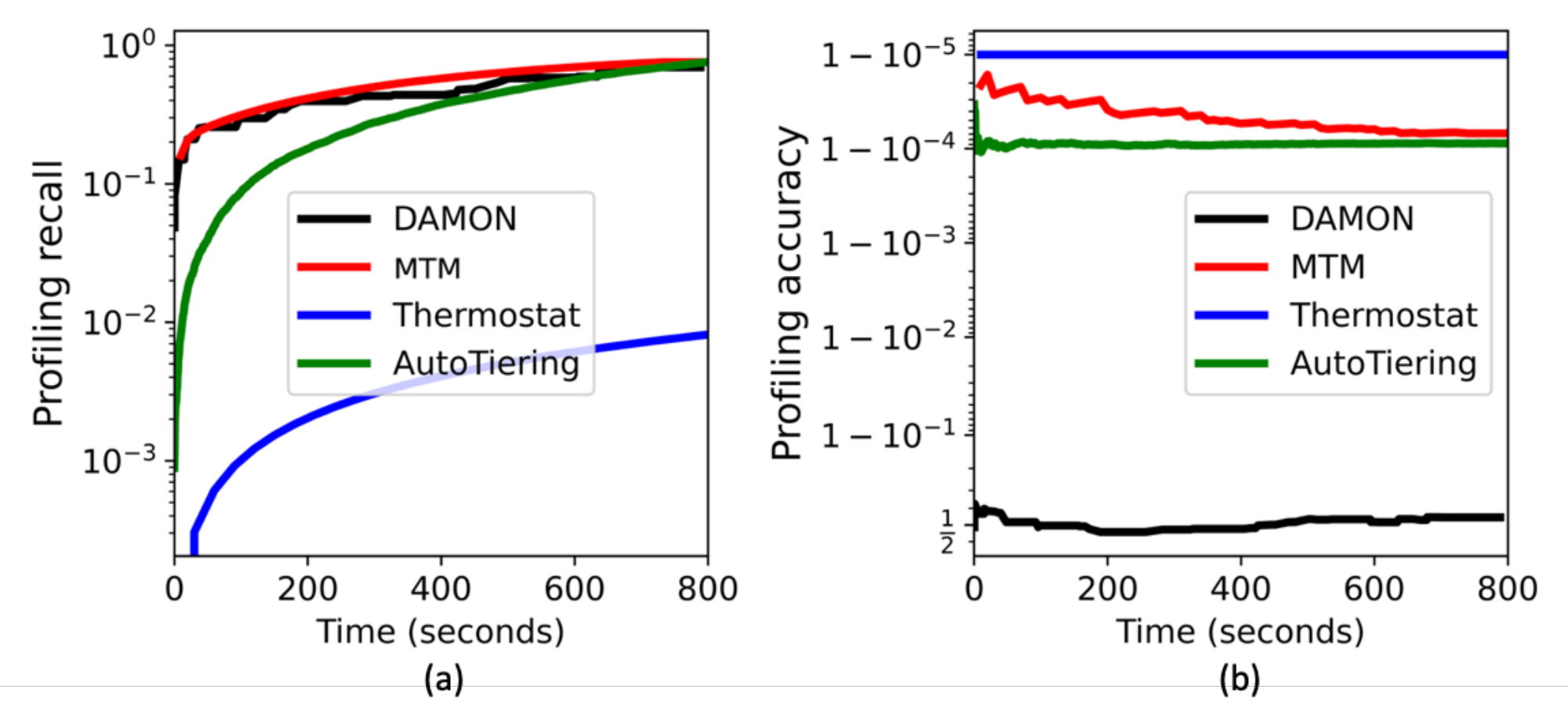}
        \vspace{-15pt}
        \caption{Comparison of different memory profiling methods in terms of their effectiveness of identifying hot pages.}
        \label{fig:motivation}
\end{figure}

We compare the profiling methods in these works with \name by running the GUPS~\cite{gups} benchmark with a 512GB working set on the four-tier memory system in Figure~\ref{fig:multi-tiered-mem}. We know a priori page hotness in each profiling interval during execution. Figure~\ref{fig:motivation} reports profiling $recall$ (i.e., the ratio of the number of correctly detected hot pages to the number of hot pages identified by priori knowledge) and profiling $precision$ (i.e., the ratio of the number of correctly detected hot pages to the number of total detected hot pages including mis-identified hot pages). 

Under the same profiling overhead (5\%), Thermostat and AutoTiering take long time to reach high recall (i.e., slower to identify hot pages in Figure~\ref{fig:motivation}.a),  because their randomness in page sampling and the formation of memory regions cause uncontrolled profiling quality. DAMON takes shorter time in Figure~\ref{fig:motivation}.a, but about 50\% of hot pages detected by DAMON are not hot (see Figure~\ref{fig:motivation}.b), because of its ad-hoc design of forming regions and slow response to the variance of memory access patterns. Due to low profiling quality, DAMON, Thermostat, and AutoTiering perform 15\% worse than \name.


We rethink memory profiling support for multi-tiered large memory based on the latest Linux support (i.e., DAMON). DAMON splits a process's virtual memory space into memory regions. In each profiling interval, it randomly profiles one 4KB page per memory region to capture spatial locality. The control of profiling overhead is achieved through a user-defined maximum number of memory regions for profiling. DAMON merges two neighbor memory regions if they have similar profiling results in an interval, and  \textcolor{dong}{splits  \textit{each} region into two randomly-sized regions} to improve profiling quality if fewer than half of the maximum number of regions exist.

We identify five limitations to be addressed.

\begin{enumerate}[leftmargin=*,noitemsep,topsep=0pt]
    \item The control of profiling overhead is not directly connected with the profiling mechanism (i.e., scanning PTEs), but connected with the number of memory regions and only one random page in a region is profiled. This constraint compromises profiling quality.
    
    \item Splitting memory regions is ad-hoc. \textcolor{dong}{Splitting each region into two can lead to useless profiling}  when the new regions after splitting have the similar memory access patterns.

    \item The process of forming memory regions can be slow to capture access patterns, because of the time constraint in the profiling intervals. This is problematic in large memory systems (e.g., multi-terabyte capacity) with many memory regions to split and merge.
    
    \item Temporal locality is not considered well.
    
    \item Lack of support for profiling huge pages.
\end{enumerate}

\section{Overview}
\label{sec:overview}


\name is designed to control the profiling overhead by considering the total number of PTE scans in all memory regions (Section~\ref{sec:overhead_control}). Like DAMON in Linux, \name partitions the virtual address space of a process into memory regions for profiling and dynamically merges and splits them. However, \name has the freedom to perform PTE scan multiple times for a page or multiple pages in a memory region in a profiling interval (Section~\ref{sec:page_sampling}). Having such flexibility provides opportunities to re-distribute sampling quotas between memory regions under a fixed profiling overhead to improve profiling quality, addressing Limitations 1 and 2. \name uses performance counter-assisted PTE scanning to quickly detect changes in access patterns and address Limitation 3 (Section~\ref{sec:hybrid_profiling}).

\name decide page migration between tiers based on a ``global view'' of all memory regions (Section~\ref{sec:migration}). 
By calculating the exponential moving average of page hotness collected from all profiling intervals, \name learns the distribution of hot memory regions in \textit{all} memory tiers, addressing Limitation 4. 




Guided by the new profiling techniques, \name introduces the ``fast promotion and slow demotion'' policy (Section~\ref{sec:where_to_migrate}). Also, when migrating pages, \name uses an asynchronous page-copy mechanism that overlaps page copying with application execution. This mechanism reduces the overhead of page copy, but come with the time cost of \textit{extra} page copy, because when a page is updated during copying, the page has to be copied again. The traditional, synchronous page-copy mechanism does not need extra page copy, but completely exposes the overhead of page copy into the critical path. Hence, \name uses a hybrid approach that takes advantage of both mechanisms, and selects one based on whether page modification happens during migration (Section~\ref{sec:mechanism}).  

To support huge pages, \name enables page profiling at the huge page level instead of 4KB-page level using the page table information; memory merging and splitting are carefully managed to be aligned with the huge page size, such that the huge page semantics is not broken (Section~\ref{sec:profile_huge_page}). Also, \name lifts the constraint in existing work that huge pages cannot move across distant memory tiers and support huge-page migration based on the existing support of 4KB-page migration (Section~\ref{sec:migration_huge_page}).



In summary, \name has (1) an adaptive profiling mechanism with high profiling quality and constrained overhead; (2) a page-migration strategy using a global view to make informed decisions; and (3) a page-migration mechanism adapting data copy schemes based on page access patterns.

\section{Adaptive Memory Profiling}
\label{sec:profiling}
\name tracks page accesses using PTE reserved bits and PTE scan. Each PTE maintains an access bit, indicating its access status. The access bit is initially set to 0, but changed to 1 by the memory management unit (MMU) when the corresponding page is accessed. By repeatedly scanning PTE to check the value of the access bit and resetting the access bit to 0 if found to be 1, page accesses can be monitored. This mechanism is commonly used in Linux and existing works~\cite{Hirofuchi:2016:RHV:2987550.2987570,middleware19_profiling}. 


Scanning all PTEs to track memory accesses to each page is prohibitively expensive for large memory. For example, scanning a five-level page table for 1.5 TB memory in 2MB pages on an Optane-based platform (hardware details in Table~\ref{tab:hardware}) with helper threads takes more than one second -- 
infeasible to capture workload behaviors online. Thus, page sampling is often used to avoid such high overhead. However, large memory systems have large numbers of pages and the profiling quality with unguided, random sampling~\cite{Agarwal:2017:TAP:3037697.3037706,intel_mem_optimizer,memory_tiering,atc21_autotiering,middleware19_profiling} could lead to poor performance. \name systematically forms memory regions to address this problem.

\subsection{Formation of Memory Regions}

A memory region is a contiguous address space mapped by a last-level page directory entry (PDE) by default. This indicates that in a typical five-level page table, the memory region has a default size of 2MB. During program execution, whenever a last-level PDE is set as valid by the OS, the corresponding memory region is subject to profiling.

\textbf{Multiple scans of PTEs.} In a profiling interval, the access bit of the PTE corresponding to a sampled page is scanned multiple times. The total number of scans per PTE per profiling interval is subject to a constant, $num\_{scans}$.


We use the multi-scan method, instead of single-scan to reduce skewness of profiling results. 
In a profiling interval, a single-scan method can only detect whether a page is accessed or not, but cannot accurately capture the number of accesses. Although aggregating memory accesses across multiple profiling intervals could alleviate this problem, the skewness of profiling results can be accumulated over time (see Section~\ref{sec:which_mem}), leading to sub-optimal migration decisions. Using the multi-scan method avoids this problem. 

At the end of a profiling interval, the average number of accesses to all sampled pages in a memory region is used to indicate the \textit{hotness} of that memory region. Based on the results, \name may merge or split memory regions. Note that the formation of memory regions through merging and splitting is based on ``logical'' memory regions and there is no change to PTE during the formation.

\textbf{Merge memory regions.}
\name actively looks for opportunities to merge contiguous memory regions at the end of a profiling interval. Two contiguous regions are merged, if their difference in the hotness in the most recent profiling interval is smaller than a threshold $\tau_1$. 


\textbf{Split a memory region.}
\name checks whether a memory region should be split to keep pages in a region to have similar hotness. When the maximum difference in the number of accesses among all sampled pages in a region exceeds a threshold $\tau_2$, the region is split into two equally-sized ones. 

\textbf{Selection of $\tau_1$ and $\tau_2$.}
$\tau_1$ and $\tau_2$ define the minimum and maximum differences in the number of memory accesses among sampled pages in a memory region. $\tau_1$ and $\tau_2$ fall into [0, $num\_scans$]. To avoid frequent merging/splitting and balance between them, $\tau_1$ and $\tau_2$ evenly split the range of [0,  $num\_scans$], i.e., $\tau_1 = 1/3 * num\_scans$ and $\tau_2 = 2/3 * num\_scans$ by default. $\tau_1$ can be dynamically fine-tuned to enforce the limit on profiling overhead, discussed in Section~\ref{sec:overhead_control}.



\subsection{Adaptive Page Sampling}
\label{sec:page_sampling}


Because the profiling overhead control is decoupled from the number of memory regions and relies on PTE scans, \name can adapt the number of sampled pages in a region to improve profiling quality. Since each sampled page has the same number of PTE scans per profiling interval, the control of PTE scans is implemented by the control of page samples (see  Section~\ref{sec:overhead_control}).


\textbf{Initial page sampling.} \name differentiates the slowest tier from other tiers in each profiling interval. The slowest tier relies on performance counters to identify memory regions with memory accesses. Only those memory regions are subject to be profiled with PTE scan-based profiling. Each memory region in the slowest memory tier for profiling has only one page profiled. This page is the one captured by performance counters. 
All other memory tiers profile all memory regions and each region initially has a random page profiled.

\textbf{After merging two memory regions}, the total number of page samples in the two regions is reduced by half under the constraint that the new region has at least one sample. This reduction saves the profiling overhead for the two regions, and allows other memory regions to have more samples without exceeding the overhead constraint.

The saved page sample quota after merging memory regions is re-distributed to other memory regions. First, \name distributes sample quota to the memory regions whose hotness indication shows the largest variance in the last two profiling intervals among all memory regions. Having a large variance of hotness indication in two profiling intervals indicates the memory access pattern change. Adding more page samples for profiling in this case is useful to improve profiling quality.

To efficiently find memory regions with the largest variance of hotness indication among all regions, \name keeps track of top-five largest variances and the corresponding regions when analyzing profiling results. \textcolor{check}{We choose ``five'' empirically to make it lightweight}. Whenever a new profiling result for a region is available, \name checks the top-five records and updates them if needed. After the merging, the saved page-sample quota is  re-distributed to those top-five regions.



\textbf{After splitting a memory region} into two new regions, the page sample quota in the original region is evenly split into the two regions. Therefore, splitting does not change the number of total PTE scans. Splitting memory regions brings two benefits. First, the hotness indication, which is the \textit{average} number of accesses to all sampled pages in a memory region, provides better indication of memory accesses to the new, smaller memory regions, hence providing better guidance on page migration. Second, migration is more effective, because using the smaller memory region avoids unnecessary data movement coming with the larger region.



\subsection{Profiling Overhead Control}
\label{sec:overhead_control}
\name supports the user to define a profiling overhead constraint. \name respects this overhead constraint while maximizing profiling quality, by dynamically changing the number of memory regions and distributing page-sample quotas between the regions. The overhead constraint is a percentage of program execution time without profiling and migration. For example, in our evaluation section, this overhead constraint is 5\%.
Given the length of a profiling interval ($t_{mi}$), profiling overhead constraint, overhead of scanning one PTE ($one\_scan\_overhead$), and the number of scans per PTE ($num\_scans$), the total number of page samples in all memory regions that can be profiled in a profiling interval, denoted as $num\_ps$, is calculated in Equation~\ref{eq:profiling_overhead_control}.


\vspace{-10pt}
\begin{equation}
\label{eq:profiling_overhead_control}
    num\_{ps} = \frac{t_{mi} \times profiling\_overhead\_constraint }{one\_scan\_overhead \times num\_{scans}}
\end{equation}


$t_{mi}$ can be set by the user, as in existing works~\cite{Agarwal:2017:TAP:3037697.3037706, Hirofuchi:2016:RHV:2987550.2987570,middleware19_profiling}. $one\_scan\_overhead$ is measured offline.
As \name merges or splits memory regions, the total number of page samples in all memory regions remains equal to $num\_ps$ to respect the profiling overhead constraint. 



The total number of memory regions needs to be smaller than $num\_ps$ so that each memory region has at least one page sample. \textcolor{check}{When the total number of memory regions is too large, \name temporarily increases $\tau_1$  
(the threshold to merge regions, and keeps $\tau_2 > \tau_1$) 
to merge memory regions more aggressively. $\tau_1$ is gradually increased across profiling intervals, until the number of memory regions is no larger than $num\_ps$, and then $\tau_1$ is reset to the original value.}





Another approach to enforce the profiling overhead constraint is to change the number of scans per page sample (i.e., $num\_scans$). However, we do not use this approach,  
because of its significant impact on profiling quality. Changing $num\_scans$ leads to a change of profiling results in \textit{all} memory regions. For example, in our evaluation, when changing $num\_scans$ from 2 to 3, \name changes the migration decision for at least 20\% of memory regions. We set $num\_scans$ as a constant ``3''. Our empirical study shows that using a value larger than 3 leads to no obvious change (less than 5\%) in the migration decision.


\textbf{Memory consumption overhead.} For each memory region, \name stores its hotness as an integer. Given a terabyte-scale memory, this yields an overhead in hundreds of MBs. In our 1.5 TB memory platform, the memory overhead to store profiling results is no larger than 600MB.

\subsection{Support of Huge Pages}
\label{sec:profile_huge_page}
When \name selects a page for profiling, this selection is huge page-aware. In particular, in the beginning of the  profiling interval, \name checks whether the selected page for profiling is a huge page or not by checking its PTE. If it is, then any access to the huge page is captured by scanning its PTE. This method is different from the existing huge page-aware solution (Thermostat). Thermostat randomly selects a basic 4KB page out of the huge page to approximate the number of memory accesses to the huge page, losing profiling quality. Also, if the huge page is \texttt{mprotected}, 4KB-based profiling in Thermostat cannot happen. \name does not have these limitations.

The formation of memory regions is also huge page-aware. When splitting a memory region, \name checks if the split happens in the middle of a huge page. If so, the split will be slightly adjusted to be aligned with the huge page boundary. Without such handling, a huge page can be profiled in two subregions, wasting PTE scans, and also suffer from conflicting migration decisions. Furthermore, the two subregions after the adjustment may not be equally-sized, increasing the risk of memory fragmentation after migration. However, in practice, given the large size of the memory region (at the scale of MBs), the size difference between the two subregions is small (often less than 10KB) without increasing memory fragmentation.


\subsection{Performance Counter-Assisted PTE Scanning}
\label{sec:hybrid_profiling}
After the initial page sampling (Section~\ref{sec:page_sampling}), the slowest memory tier continues to use performance counters to assist PTE scanning. In particular, at the beginning of a profiling interval, the performance counters are enabled for a short time (10\% of the profiling internal) to collect memory accesses and identify huge page-sized memory regions where the memory accesses happen. Those memory regions are compared with the existing memory regions in the slowest memory tier. If they are the same, then the existing regions continue to use PTE scans. If not, then the memory regions identified by the performance counters are subject to be profiled, and the existing memory regions not identified by the performance counters are not profiled  \textcolor{dong}{in the current profiling interval.} 

Compared with DAMON, using performance counters can save multiple profiling intervals to identify hot pages, because DAMON uses a time-interval-based approach to capture hot pages by chance, while \name uses an event-driven approach: once a memory region is accessed, it is immediately subject for high-quality profiling to confirm its hotness. \textcolor{check}{Different from Dancing~\cite{ipdps21:dancing} that also uses the hybrid performance counter/PTE scan, \name selectively apply the hybrid approach  \textcolor{dong}{(i.e. only in the slowest memory tier with the largest capacity)} and the interaction between the counters and PTE scans are driven by the needs of fast responses (not gaining better visibility to memory accesses as Dancing).} Note that using performance counters alone without PTE scan is not enough to provide high-quality profiling, shown in our comparison with HeMem (Section~\ref{sec:eval_overall_perf}), because its full randomness misses hot pages.

\section{Page Migration Strategy}
\label{sec:migration}
A page migration strategy decides (1) which memory regions to migrate, and (2) given a region to migrate, which memory tier to migrate. 




\subsection{Which Memory Region to Migrate?}
\label{sec:which_mem}
At the end of a profiling interval, \name promotes some memory regions to the fastest tier, and the total size of migrated regions is a constant $N$ (N=200MB in our evaluation). This is similar to the existing works~\cite{intel_mem_optimizer, memory_tiering, atc21_autotiering,ATC20_leap} that periodically migrates a fixed number of pages. If there is no enough free space in the fastest tier, some pages in the fastest tier need to be demoted first to the slower tiers (see Section~\ref{sec:where_to_migrate}).


\textbf{Select memory regions for promotion.}
The goal of promotion is to place recent frequently accessed pages into faster tiers. \name's migration decision is holistic -- considering \textit{all} memory regions together regardless of which tiers those memory regions are currently in. \name uses time-consecutive profiling results based on the exponential moving average (EMA) of hotness indication collected from all profiling intervals. Given a sequence of data points, EMA places a greater weight and significance on the most recent data points. Thus, \name using EMA considers temporal locality in migration decision and avoids page migration due to the bursty memory access pattern in one profiling interval. 

We define EMA of hotness indication as follows. Assume that $HI_i$ is the hotness indication collected at the profiling interval $i$ for a memory region, and the EMA of hotness indication for that memory region at $i$, denoted as $WHI_i$, is defined in Equation~\ref{eq:whi}. This equation is a recursive formulation including $WHI_i$ and $WHI_{i-1}$ from the prior interval $i-1$.

\vspace{-10pt}
\begin{equation}
\label{eq:whi}
\small
    WHI_i = \alpha \times HI + (1-\alpha) \times WHI_{i-1}
\end{equation}




$\alpha$ indicates the importance of historical information in decision making. In practice, we set $\alpha$ as 0.5. There are two benefits of using EMA. First, the memory consumption is small. There is no need to store all prior profiling results. Second, the computation of EMA is lightweight. 


With the EMA of hotness indication in all memory regions, \name builds a histogram to get the distribution of EMA of all memory regions. The histogram buckets the range of EMA values, and counts how many and what memory regions fall into each bucket. Given the size of pages to migrate to the fastest memory, \name chooses those memory regions falling into the highest buckets in the histogram to migrate. Building and maintaining the histogram has low overhead: whenever the EMA of hotness indication of a memory region is available, the histogram only needs to be slightly updated accordingly.


\subsection{Where to Migrate Memory Regions?}
\label{sec:where_to_migrate}



\textbf{Promotion.} As memory regions are promoted to the fastest tier based on the histogram, it is likely that after a profiling interval, there is no region to promote because those regions falling into the highest buckets of the histogram are already in the fastest tier. In that case, memory regions in the lower buckets of the histogram are selected to promote to the second-fastest memory tier. The accumulated size of regions to migrate should always be $N$. In general, \name makes the best efforts to promote frequently accessed regions to faster tiers.

\textbf{Demotion.} When a memory tier is a destination of memory promotion but does not have enough space to accommodate memory promotion, some memory regions in that memory tier may need to be demoted to the next lower memory tier with enough memory capacity. Memory regions for demotion are selected based on the histogram -- memory regions that are in the lowest buckets of the histogram are demoted to the next lower tier. We use the above slow-demotion strategy to avoid performance loss caused by migrating pages that are still likely to be accessed in near future. 

\textbf{Handling multi-view of tiered memory.} \textcolor{dong}{If the application managed by \name has multiple threads, depending on which memory node a thread is close to, different threads can have different views on whether a memory node is fast or slow. For example, in Figure~\ref{fig:multi-tiered-mem}, a thread on processor 0 thinks DRAM 0 is a fast memory,  while a thread on processor 1 thinks DRAM 0 is a slow memory. We call this, the \textit{multi-view of tiered memory}.  The multi-view does not impact  profiling results because the page hotness is only related to the number of page accesses, no matter from which thread memory accesses come. However, the multi-view impacts page migration destination.}

\textcolor{dong}{In \name, the migration destination of a page is decided using the view of the thread that has the most accesses to that page, because enabling high performance for the most memory accesses gives the best overall performance. To support this design, during the memory profiling, \name checks where memory accesses come from. This is implemented by leveraging the hint-fault mechanism in Linux~\cite{autonuma}. A hint-fault takes 12$\times$ longer time than a PTE scan. To amortize this cost, every 12 PTE scans, 
\name turns on the hint-fault mechanism to capture one memory access, and includes the amortized cost into $one\_scan\_overhead$ in Equation~\ref{eq:profiling_overhead_control}}. 



\section{Adaptive Migration Mechanism}
\label{sec:mechanism}

\begin{figure}[tb]
\centering
\includegraphics[width=1\columnwidth]{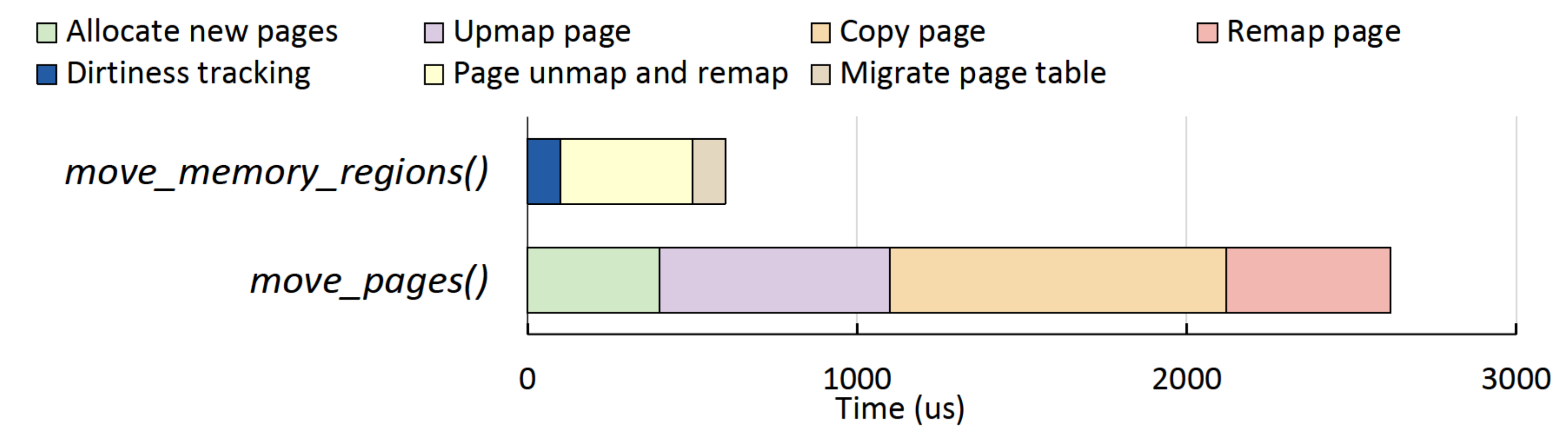}
\vspace{-15pt}
\caption{Performance breakdown for migration mechanisms.}
\label{fig:movemr_breakdown}
\end{figure}


\subsection{Performance Analysis of Page Migration Mechanism}
Linux provides an API, \textit{move\_pages()}, for a privileged process to move a group of 4KB pages from a source memory node (or tier) to a target memory node (or tier). There are four steps in \textit{move\_pages()}, and they are performed sequentially: (1) allocate new pages in the target memory node; (2) unmap pages to migrate (including invalidating PTE); (3) copy pages from source to target memory node; (4) map new pages (including updating PTE).

Figure~\ref{fig:movemr_breakdown} shows the performance of migrating a 2MB memory region from the fastest tier to the slowest tier on the Optane-based platform. Copying pages is the most time-consuming step, taking 40\% of total time. One reason for the high overhead is that \textit{move\_pages()} moves 4KB-sized pages sequentially. Although recent work~\cite{Yan:2019:NPM:3297858.3304024} enables multi-threaded page copy to fully utilize memory bandwidth, it is still a bottleneck especially when moving a large memory region.

\subsection{Adaptive Page Migration Schemes}

\textbf{Asynchronous page copy.} We introduce an asynchronous page copy mechanism to reduce page copy overhead. 
In the asynchronous page copy, the thread that triggers migration (named \textit{main thread}) launches one or more helper threads to run the steps (1) and (3); the main thread runs the steps (2) and (4), and then waits for the helper thread(s) to join. With the asynchronous page copy, it is possible that copying a page happens before invalidating its PTE but the page is modified in the source memory tier after copying the page. In such cases, the page must be copied again to update its copy in the target tier, which is costly. We introduce an adaptive page-migration mechanism to address this limitation.

\textbf{Adaptive page migration.} For read-intensive pages, the asynchronous page copy brings performance benefit. However, for write-intensive pages, due to repeated data copy, it is likely that the asynchronous page copy performs worse than the synchronous. Hence, \name chooses suitable migration mechanism based on the write intensity of pages. In particular, \name uses the asynchronous page copy by default. But whenever any page in the memory region for migration is written after the asynchronous page copy starts, \name switches to the synchronous page copy. To track page write, \name utilizes PTE access bits and page faults, discussed in Section~\ref{sec:impl}. 

We implement the above mechanism and introduce a new API called \textit{move\_memory\_regions()}. In this implementation, tracking page write, performing page map/unmap, and migrating PTEs are still on the critical path, but the time-consuming page copying could be performed off the critical path. Figure~\ref{fig:movemr_breakdown} presents the performance of \textit{move\_memory\_regions()} migrating 2MB memory region using the same setting as for \textit{move\_pages()}, and excludes the overhead of page copying (and page allocation in the step (1), using asynchronous page allocation). \textit{move\_memory\_regions()} is 4.37$\times$  faster than \textit{move\_pages()} in this case. Section~\ref{sec:eval_migration_mechanism} shows more results.

\subsection{Support of Huge Page Migration}
\label{sec:migration_huge_page}
\name enables direct migration of a huge page to a lower memory tier without waiting for it to be split into basic 4KB pages as AutoTiering (a solution supporting  multi-tiered memory). In particular, \name uses \textit{move\_memory\_regions()} to move a huge page as a whole. In contrast, AutoTiering puts huge pages to the (in)active lists where they are split into 4KB basic pages for profiling and then for migration as needed. 




\section{Implementation}
\label{sec:impl}



We implement the adaptive memory profiling as a kernel module that enables performance counters to guide profiling and periodically scans PTEs based on adaptive page sampling. The kernel module takes a process ID as input. For performance counters, it uses Intel processor event-based sampling mode (PEBS) to collect memory accesses into a preallocated buffer, and uses a register interrupt handler to indicate when the buffer is full. \name uses hardware events \texttt{MEM\_LOAD\_RETIRED.LOCAL\_PMM} and  \texttt{MEM\_LOAD\_RETIRED.REMOTE\_PMM} to capture memory accesses. The sampling frequency is set as 200 memory accesses as in production environment~\cite{meta_tpp}, which is sufficient to distinguish hot and code pages~\cite{sosp21_hemem}. 
In addition, profiling results from PTE scans is saved in a shared memory space, and stored as a table, where each record contains a memory region ID, hotness indication in the current profiling interval, and the EMA of hotness indication of prior intervals. The region ID is generated based on the start address of the memory region.

We implement the page management as a daemon service at the user space. The daemon service executes with the application and calls the kernel module for profiling at the beginning of each profiling interval. At the end of each profiling interval, the service reads collected profiling results from the shared memory space. With overhead control, the profiling module ensures that profiling always finishes within a profiling interval. The service then makes the migration decision and performs migration using \textit{move\_memory\_regions()}.


\textit{move\_memory\_regions()} takes the same input as Linux \textit{move\_pages()}, but implements the adaptive migration mechanism. It detects page dirtiness during the migration by setting a reserved bit in PTE to trigger a write protection fault when there is write to the memory region. Leveraging a user-space page fault handler, \textit{move\_memory\_regions()} tracks writes, and decides whether to stop the asynchronous page copy and switch to the synchronous mechanism. 

\textcolor{dong}{\name cannot manage multiple applications co-running in the same machine, because that requires coordination of page migration among applications, which is our future work.}

\section{Evaluation}
\label{sec:eval}

\textbf{Testbed.} We evaluate \name on a two-socket machine. Each socket has Intel Xeon Gold 6252 CPU (24 cores), 756GB Intel Optane DC PM and 96GB DRAM. This machine has four memory tiers (see Figure~\ref{fig:multi-tiered-mem}). We also emulate a four-tiered memory system with doubled latency and bandwidth using memhog in Linux. The emulation launched memhog instances in each memory tier to continuously inject memory access traffic. The details of evaluation platforms show in table~\ref{tab:hardware}.
Unless indicated otherwise, 
we report results on the Optane-based machine (not the emulation platform). 
We use \texttt{madvise} for THP~\cite{thp} and uses 2MB as huge page size by default, which    
is typical in large memory systems. We set the profiling overhead constraint to 5\% and the profiling interval to 10 seconds. This setting is similar to existing works and production environments~\cite{Agarwal:2017:TAP:3037697.3037706, Hirofuchi:2016:RHV:2987550.2987570,meta_tpp, middleware19_profiling}. 




\textbf{Workloads.} We use large-memory workloads 
in Table~\ref{tab:app_details}. Their memory footprints are larger than the fastest two tiers, enabling effectively evaluation on all tiers. Unless indicated otherwise, we use eight application threads per workload.

\begin{table}[t]
\caption{Hardware overview of experimental system.}
\footnotesize
\begin{tabular}{ll}
\hline
\multicolumn{2}{c}{\textbf{Optane-based Multi-tiered Memory System}}          \\ \hline
Fast Mem Local Access (1st tier)              & latency: 90ns \phantom{LLL} bw: 95 GB/s    \\
Fast Mem Remote Access (2nd tier)            &  latency: 145ns \phantom{LL}  bw: 35 GB/s \\
Slow Mem Local Access (3rd tier)              &  latency:  275ns \phantom{LL} bw: 35 GB/s\\
Slow Mem Remote Access (4th tier)              &  latency: 340ns \phantom{LL}  bw: 1 GB/s \\\hline \hline
\multicolumn{2}{c}{\textbf{Emulated Multi-tiered Memory System}}          \\ \hline
Fast Mem Local Access (1st tier)              & latency: 198ns  \phantom{LL} bw: 95 GB/s    \\
Fast Mem Remote Access (2nd tier)            &  latency: 315ns \phantom{LL}  bw: 35 GB/s \\
Slow Mem Local Access (3rd tier)              &  latency: 825ns  \phantom{LL} bw: 35 GB/s \\
Slow Mem Remote Access (4th tier)              &  latency: 1010ns \phantom{L}  bw: 1 GB/s \\\hline

\end{tabular}
\label{tab:hardware}
\vspace{-5pt}
\end{table}
\begin{table}[t]
\caption{Workloads for evaluation.}
\scriptsize	
\begin{tabular}{p{1.1 cm}|p{4.6cm}|p{0.55 cm}|p{0.5 cm}}
\hline
Workloads & Descriptions & Mem  & R/W \\ \hline
GUPS~\cite{gups} &  A measurement of how frequently a computer can issue updates to randomly generated memory locations.  &      512GB    &    1:1        \\  \hline
VoltDB~\cite{voltdb}  &  A commercial in-memory database with TPC-C~\cite{tpcc} using 5K warehouse.  &          300GB /1.2TB     &    1:1              \\ \hline
Cassandra~\cite{cassandra} &   A highly-scalable partitioned row store with YCSB~\cite{ycsb} (using update-heavy benchmark A).            &      400GB    &    1:1          \\ \hline
BFS~\cite{peng2018graphphi}&    A parallel implementation of graph traversing and searching on a graph with 0.9B nodes and 14B edges.  &          525GB          &    read-only              \\ \hline
SSSP~\cite{peng2018graphphi}&  A 
parallel implementation of finding the shorted path between two vertices on a graph with 0.9B nodes and 14B edges.           &    525GB            &        read-only          \\ \hline
Spark~\cite{spark} &  A spark program running the TeraSort benchmark~\cite{spark_terasort}. &       350GB             &      1:1            \\ \hline
\end{tabular}
\label{tab:app_details}
\end{table}

\textbf{Baselines.} Eight state-of-the-art solutions are used. 
\begin{itemize}[noitemsep]
\item \textit{Hardware-managed memory caching (HMC)} uses the fast memories as hardware-managed cache for slow memories. We use Memory Mode in Optane.

\item \textit{First-touch NUMA} is a common allocation policy. It allocates pages in a memory tier close to the running task that first touches the pages. It does not migrate pages. 



\item \textcolor{dong}{\textit{Tiered-AutoNUMA}~\cite{tiered-autonuma} in Linux and \textit{AutoTiering}~\cite{atc21_autotiering}.}

\item \textit{HeMem}~\cite{sosp21_hemem} is for two-tiered systems. HeMem leverages performance counters alone to find hot pages. 

\item \textit{Thermostat}~\cite{Agarwal:2017:TAP:3037697.3037706} is for two-tiered systems. It allocates all pages in the fast tier and selectively moves them to the slow tier. It cannot support applications with footprint larger than the fast tier. Thus, we do not evaluate its page migration but only evaluate its profiling method. 


\item \textit{DAMON}~\cite{middleware19_profiling, damon} is a Linux profiling tool at the memory region granularity. We use it to evaluate profiling quality.

\item \textit{Nimble}~\cite{Yan:2019:NPM:3297858.3304024} is a page migration mechanism using bi-direction page copy and parallel page copy. \name includes the techniques in Nimble but adds adaptive migration. We use Nimble to evaluate our page migration mechanism.

\end{itemize}

\subsection{Overall Performance}
\label{sec:eval_overall_perf}

\begin{figure}[tb]
\centering
\includegraphics[width=0.9\columnwidth]{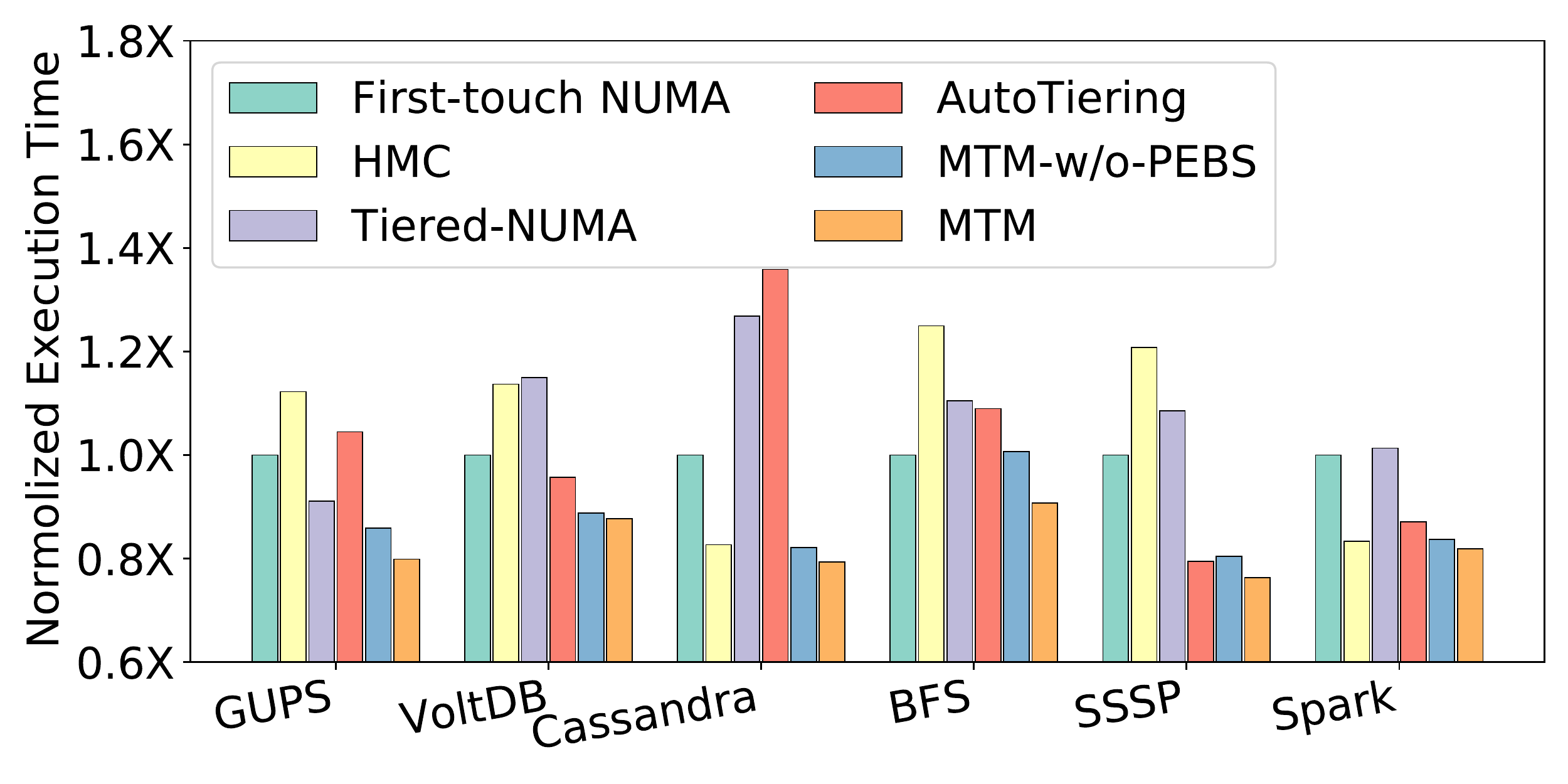}
\vspace{-10pt}
\caption{Overall performance (normalized to \textcolor{dong}{first-touch NUMA}'s) using existing solutions and \name on Optane.} 
\label{fig:overall_optane}
\end{figure}

\textbf{Optane-based multi-tiered system.}
Figure~\ref{fig:overall_optane} shows that \name outperforms all baselines. 
We have five observations. 

(1) \name outperforms HMC by up to 40\% (avg. 19\%). HMC incurs write amplification when cache misses occur~\cite{9408179}, causing unnecessary data movement and low performance.


(2) \name outperforms first-touch NUMA by up to 24\% (avg. 17\%). Without page migration, first-touch NUMA outperforms HMC on VoltDB and BFS, and outperforms tiered-AutoNUMA on Cassandra and BFS, indicating that page migration is not always helpful. HMC performs worse because of unnecessary page movement. tiered-AutoNUMA has worse performance because it cannot effectively identify hot pages. 



(3) \name outperforms tiered-AutoNUMA by up to 37\% (avg. 23\%). Tiered-AutoNUMA provides page migration across all memory tiers. However, tiered-AutoNUMA promotes hot pages slowly, 
which wastes the opportunity to accelerate application execution in fast memory. 


(4) \name outperforms AutoTiering by up to 42\% (avg. 17\%). AutoTiering uses random sampling and opportunistic demotion, failing to effectively identify pages for migration.

(5) \name outperforms the solution of disabling performance counter for profiling (MTM-w/o-PEBS in the figure) by up to 10.6\%.  
Lack of efficient methods to detect hot regions in the slowest memory tier, MTM-w/o-PEBS fails to provide high quality profiling result to guide page placement.

\begin{figure}[t!]
    \centering
        \includegraphics[width=0.9\columnwidth]{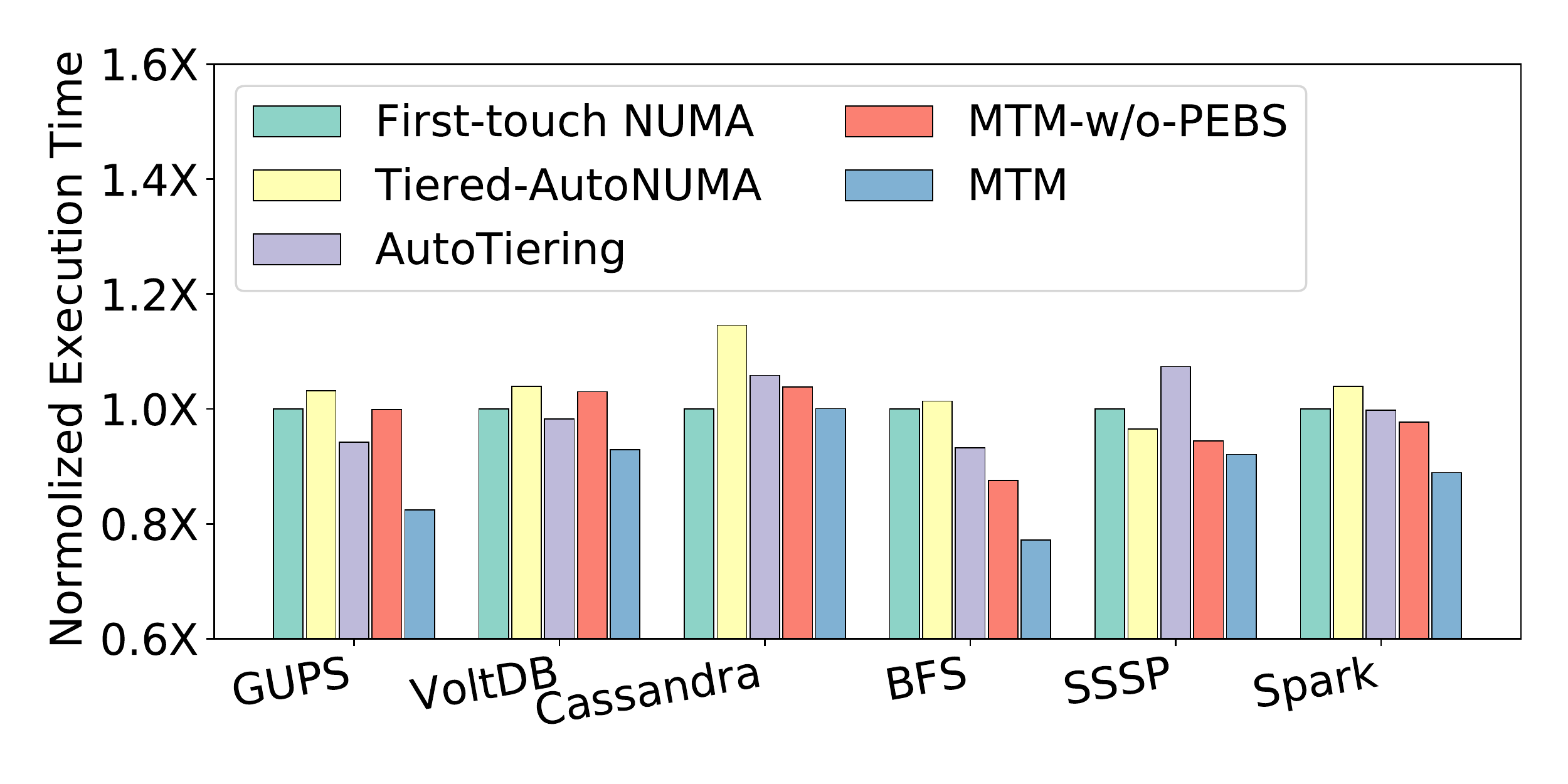}
        \vspace{-10pt}
        \caption{Performance comparison between existing solutions and \name on the emulated multi-tiered memory system. The performance is normalized by First-touch NUMA performance.}
        \label{fig:perform_emulated}
\end{figure}

\textbf{Emulated multi-tiered memory system.} Figure~\ref{fig:perform_emulated} shows that on the emulated platform, \name maintains the same performance trend as on the Optane-based testbed -- it outperforms first-touch NUMA, tiered-AutoNUMA, AutoTiering and MTM-w/o-PEBS by 19\%, 26\%, 17\%, and 14\% respectively. Note that HMC is not evaluated because when HMC is used, fast memory tiers are hidden from software and we cannot inject latency into fast memory. 

\begin{figure*}[t!]
    \centering
        \includegraphics[width=2\columnwidth]{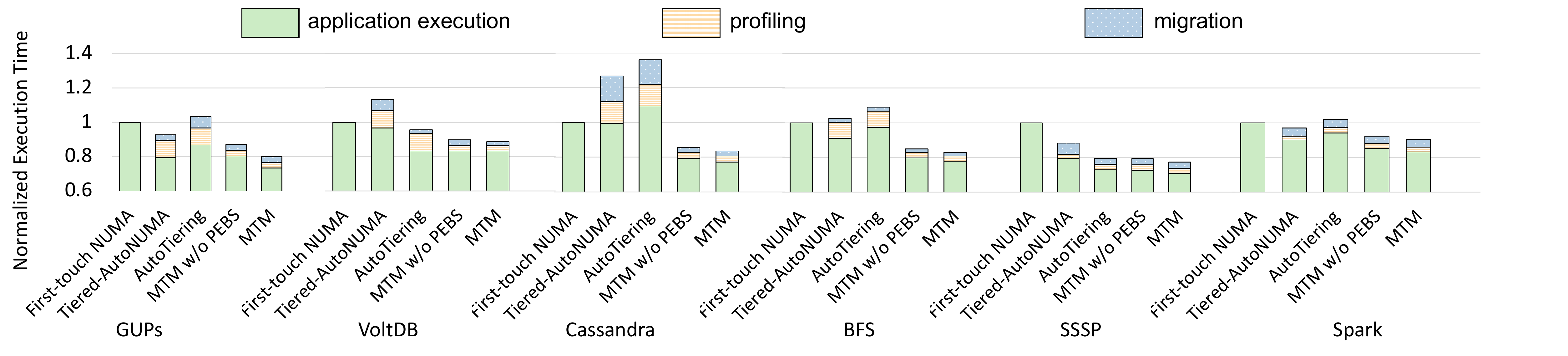}
        \vspace{-5pt}
        \caption {Breakdown of application execution time. Each bar is normalized by the execution time of first-touch NUMA. }
        \vspace{-5pt}
        \label{fig:perform_breakdown}
\end{figure*}

\textbf{Performance breakdown.}
We show the breakdown into application execution time, migration time, and profiling time in Figure~\ref{fig:perform_breakdown}. The migration time is the overhead exposed on critical path, excluding asynchronous page copying time. We only compare tiered-AutoNUMA, MTM-w/o-PEBS and \name because they are the only solutions that can leverage all four memory tiers for migration. We add first-touch NUMA as a performance baseline because evaluated solutions use first-touch NUMA for memory allocation. In all cases, the profiling overhead falls within the profiling overhead constraint.



With tiered-AutoNUMA, the time reduction is lower or equal to the overhead of profiling and migration (see VoltDB and Cassandra). Hence, they perform worse than first-touch NUMA without page migration. Compared to tiered-AutoNUMA, \name spends similar time in profiling but 3.5$\times$ faster in migration, reducing the execution time by 21\% on average. Compared to AutoTiering, \name spends similar time in profiling but 25\% faster in migration, and reduces the execution time by 19\% on average. 

\begin{figure}
    \centering
    \includegraphics[width=0.85\linewidth]{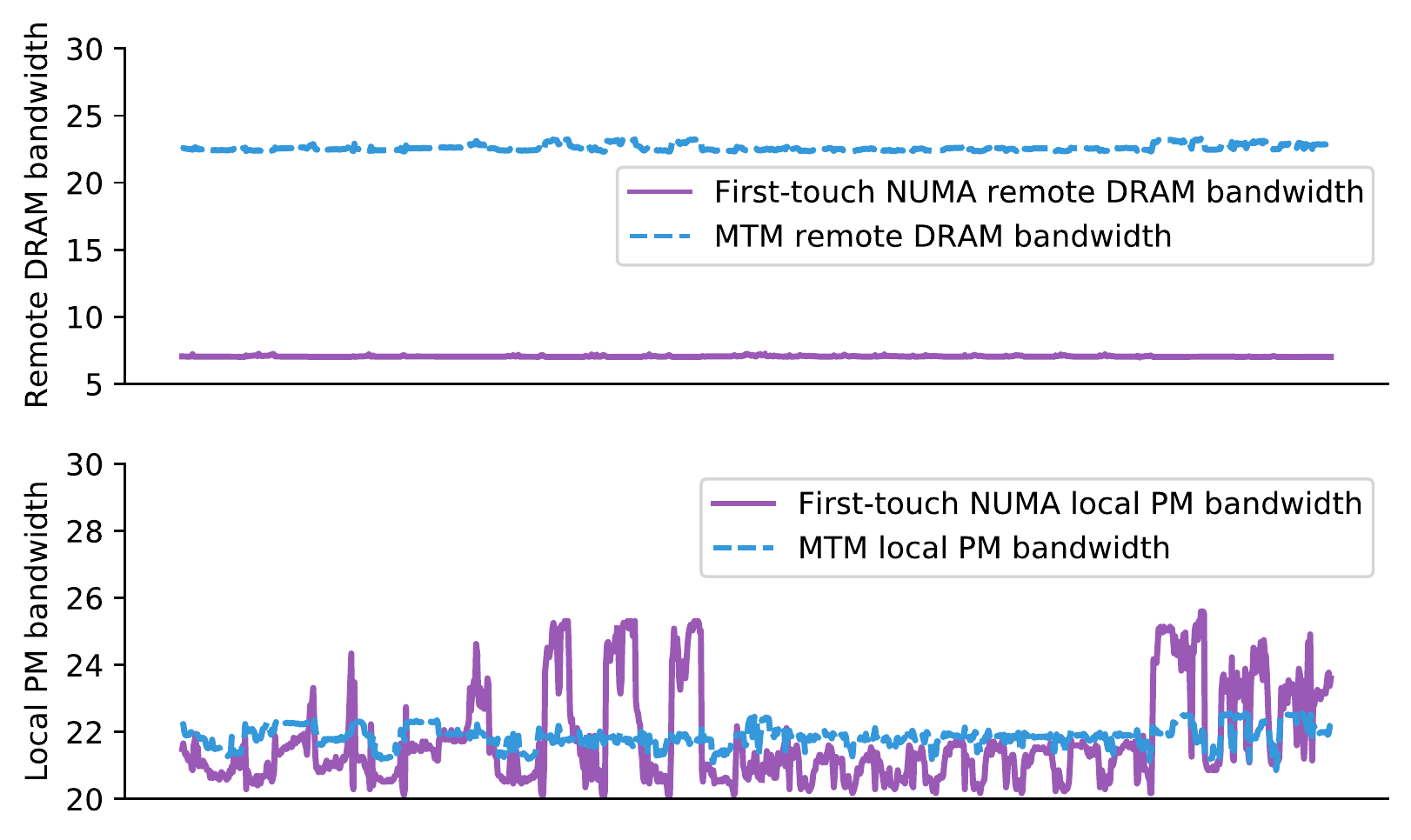}
    \vspace{-5pt}
  \caption{\textcolor{revision}{Memory bandwidth utilization with \name and NUMA first-touch}}\label{fig:bandwidth}
  \vspace{-5pt}
\end{figure}

\textcolor{revision}{\textbf{Memory bandwidth usage of \name}. We study the impact of memory bandwidth when using \name. We measure the memory bandwidth with Intel PCM~\cite{intel_pcm} and compare with NUMA first-touch, which does not have page migration. With NUMA first-touch, memory pages are allocated in local DRAM first. Once the local DRAM is full, the remaining pages are allocated in the order of local PM (tier3), remote DRAM (tier2) and remote PM (tier4). Due to space limitation, Figure~\ref{fig:bandwidth} only shows the bandwidth comparison on remote DRAM (tier2) and local PM (tier3). NUMA first-touch cannot utilize memory bandwidth offered by remote DRAM, while accesses pages on local PM, leading to under-utilization of remote DRAM. \name is able to utilize memory bandwidth from all tiers.} 


\begin{figure}
    \centering
    \includegraphics[width=0.85\linewidth]{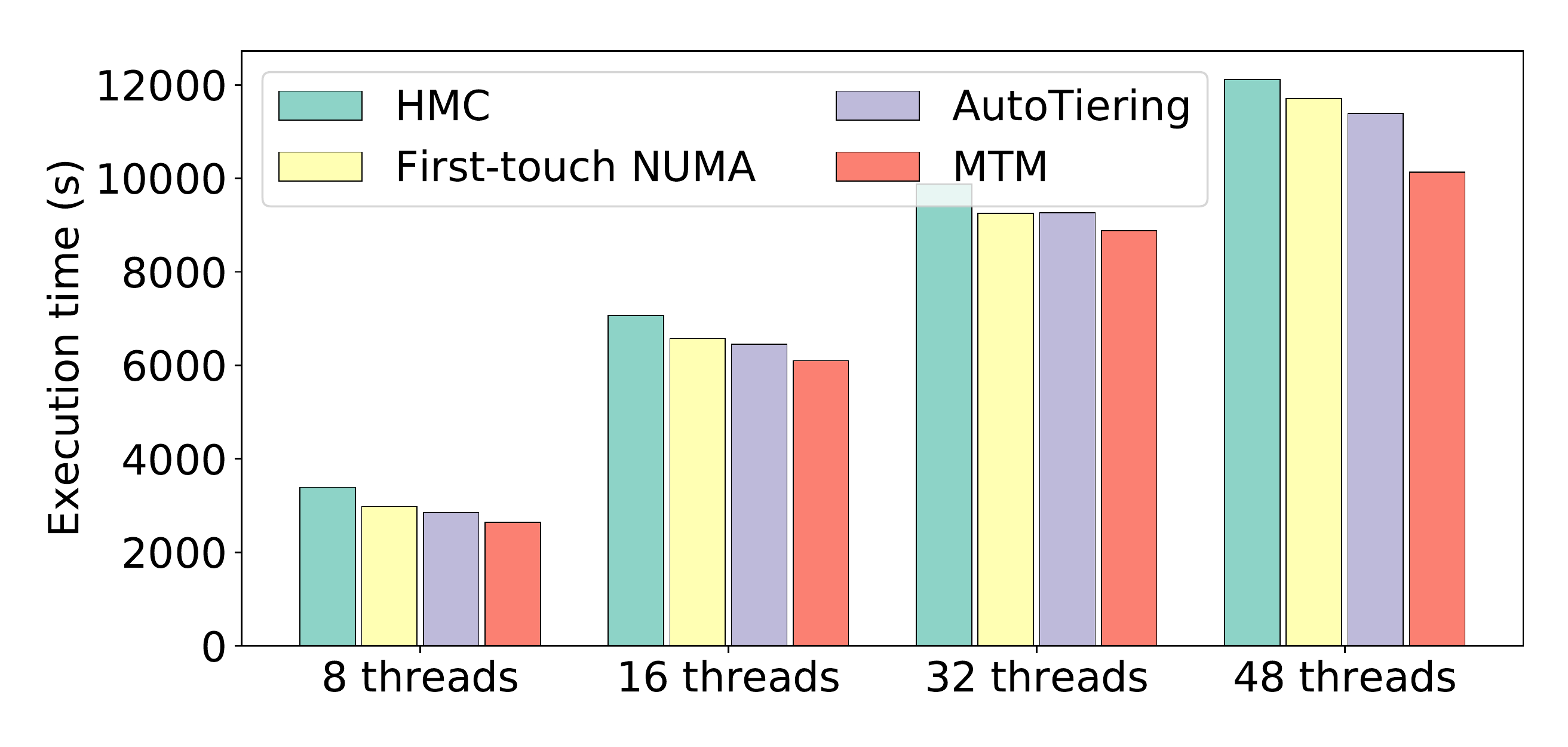}
    \vspace{-5pt}
  \caption{Execution time of VoltDB with different number of client threads.}\label{fig:scalibility}
  \vspace{-5pt}
\end{figure}

\textbf{Scalability of \name}. We evaluate the scalability of \name with VoltDB by increasing  the number of clients. We see the memory consumption increases from 300GB to 1.2TB. We compare the performance of \name, HMC, first-touch NUMA, and AutoTiering. 
We evaluate AutoTiering since it has the second-best performance among all we evaluate. Figure~\ref{fig:scalibility} shows that \name consistently outperforms HMC, first-touch NUMA, and AutoTiering by 19\%, 10\%, and 8\% on average.  

\begin{figure}
    \centering
    \includegraphics[width=0.9\linewidth]{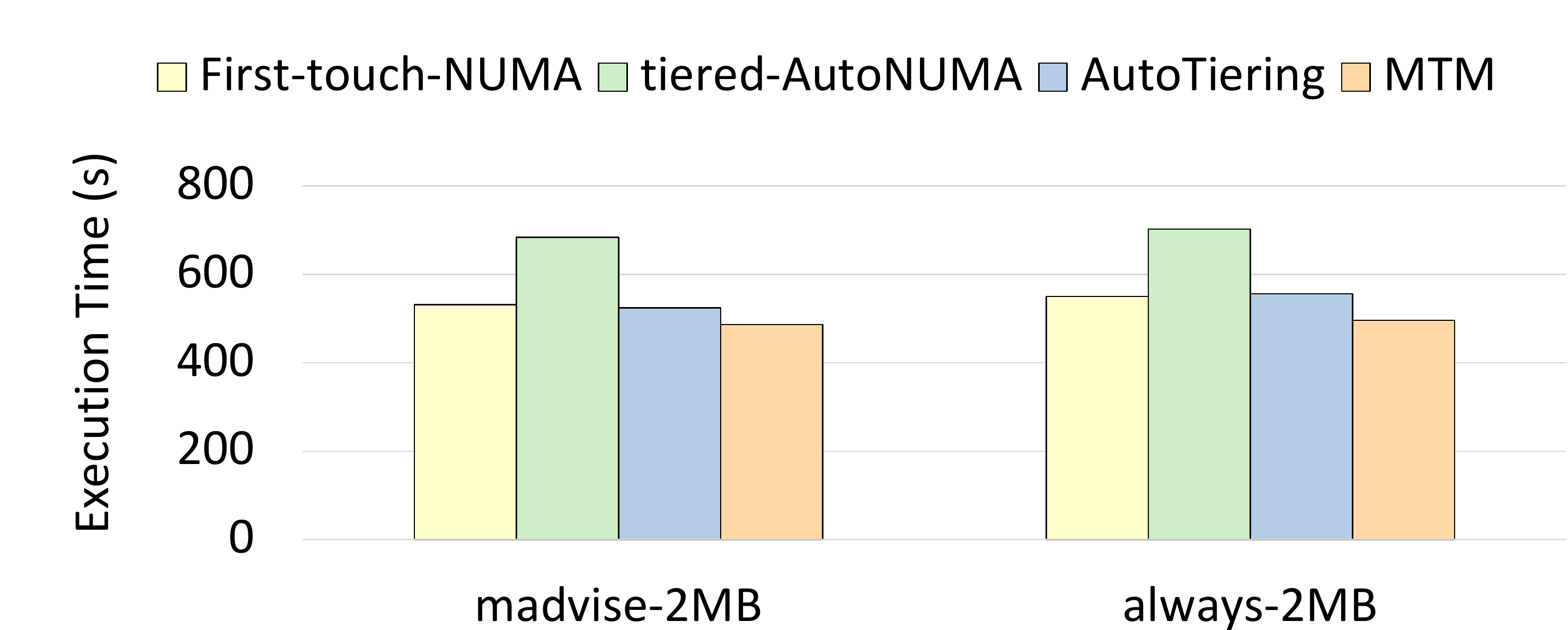}\caption{SSSP with \name and different THP configurations.}
    \label{fig:thp}
\end{figure}

\textbf{Evaluation with different THP settings.} Figure~\ref{fig:thp} shows the results using \texttt{madvise} and \texttt{always} as THP setting. We evaluate SSSP, because it has the largest memory consumption among all evaluated applications. The results confirm that \name consistently outperforms other solutions.

\subsection{Adaptive Profiling in \name}
\subsubsection{Statistic of Profiling}\hfill

\textbf{Number of memory accesses.} We count the number of memory accesses in each memory tier when running VoltDB. 
We only report the results for tiered-AutoNUMA, MTM-w/o-PEBS, and \name because only they can leverage all four memory tiers for migration. We use Intel Processor Counter Monitor~\cite{intel_pcm} to count the number of memory accesses and exclude memory accesses caused by page migration. This counting method allows us to evaluate the number of memory accesses from the application (not from page migration). Table~\ref{tab:traffic} shows the results \textcolor{dong}{where there are eight VoltDB clients residing in one processor, and the view of tiered memory is defined from their view.} Table~\ref{tab:traffic} shows that with \name, the number of memory accesses in the fastest memory tier (tier 1) is 20\% and 14\% more than with tiered-AutoNUMA and AutoTiering. This indicates that \name effectively migrates frequently accessed pages to the fast tier for high performance.

\begin{table}[]
\vspace{-10pt}
\caption{The number of memory accesses when using VoltDB.}
\footnotesize
\label{tab:traffic}
\centering
\begin{tabular}{c|c|c|c|c}
\hline
\scriptsize{\begin{tabular}[c]{@{}l@{}}\# of memory\\  accesses\end{tabular}} & \scriptsize{Tiered-AutoNUMA} & \scriptsize{AutoTiering} &\scriptsize{MTM-w/o-PEBS} & \scriptsize{\name} \\ \hline \hline
tier 1          &      248M   &  258M    &   295M   & 293M \\ \hline
tier 2           &       15M &  34M     &  198K	   & 220K  \\ \hline
tier 3 &      60M   &   30M   &   9M   & 10M \\ \hline
tier 4    &       704K  &   2.5M  &   92K  &  205K  \\ \hline
\end{tabular}
\vspace{-5pt}
\end{table}

\begin{table}[]
\caption{Statistics of forming memory regions using \name. ``MR'' and ``PI'' stand for memory regions and profiling interval.} 
\vspace{-5pt}
\footnotesize 
\label{tab:mem_regions}
\centering
\resizebox{0.8\columnwidth}{!}{%
\begin{tabular}{|c|c|c|c|c|}
\hline
          & \# of PI & \begin{tabular}[c]{@{}c@{}}avg \# of MR \\ merged in a PI\end{tabular} & \begin{tabular}[c]{@{}c@{}}avg \# of MR \\ splited in a PI\end{tabular} & \begin{tabular}[c]{@{}c@{}}avg \# of MR \\ in a PI\end{tabular} \\ \hline
GUPS      & 1000                     & 26.5                                  & 20.4                                   & 2410                            \\ \hline
VoltDB    & 800                      & 53.2                                  & 50.6                                   & 1274                            \\ \hline
Cassandra & 1600                     & 42.5                                  & 63.2                                   & 1073                            \\ \hline
BFS       & 120                      & 16.7                                  & 17.3                                   & 2574                            \\ \hline
SSSP      & 360                      & 21.3                                  & 18.2                                   & 2492                            \\ \hline
Spark     & 800                      & 35.9                                  & 32.5                                   & 1852                            \\ \hline
\end{tabular}
}%
\end{table}

\textbf{Statistics of memory regions.} On average, the number of memory regions merged and split in a profiling interval is 3.4\% of all memory regions, as reported in Table~\ref{tab:mem_regions}. Compared with DAMON, \name has less merging/splitting because of performance counter guidance and effective formation of memory regions. For example, GUPS with \name has 12\% less merging and 32\% less splitting than with DAMON.

\subsubsection{Effectiveness of Adaptive Profiling}\hfill
\label{sec:eval_profiling}

\textbf{Comparison with tiered-AutoNUMA and Thermostat.} We study profiling quality and overhead, and compare \name with two sampling-based profiling methods: one used in tiered-AutoNUMA and AutoTiering, and the other used in Thermostat. We use tiered-AutoNUMA and Thermostat, and replace their migration with \name's,  which excludes the impact of migration on performance, and is fair.



The profiling method in tiered-AutoNUMA randomly chooses a 256MB virtual address space in each profiling interval, and then manipulates the present bit in each PTE in the chosen address space. This method tracks page accesses by counting page faults. The profiling method in Thermostat randomly chooses a 4KB page out of each 2MB memory region for profiling. This method manipulates page protection bits in PTE and leverages protection faults to count accesses.

\begin{figure}[t!]
    \centering
        \includegraphics[width=0.95\columnwidth]{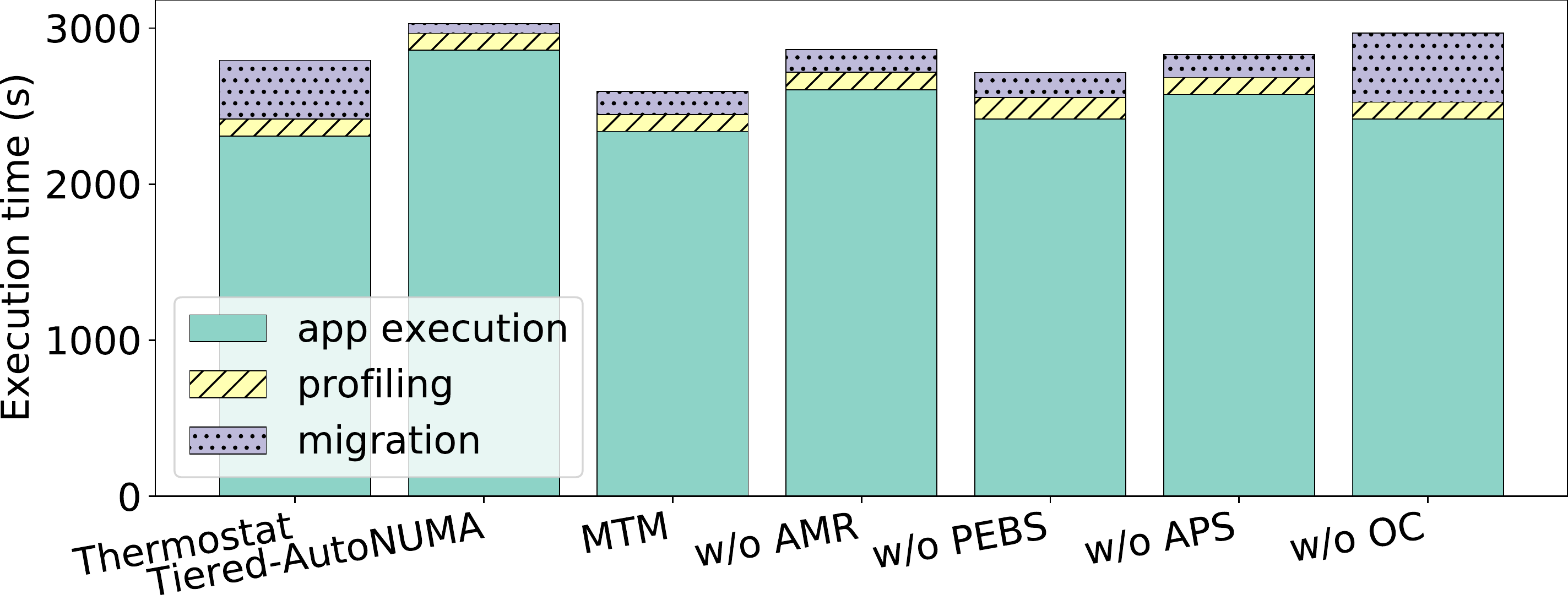}
        \caption{Evaluation of the effectiveness of adaptive memory regions (``AMR''), adaptive page sampling (``APS''), and profiling overhead control (``OC'') using VoltDB.}
        \label{fig:profiling}
\end{figure}

Figure~\ref{fig:profiling} shows that \name outperforms tiered-AutoNUMA and Thermostat by 17\% and 7\%, respectively. The profiling overhead in Thermostat is 6$\times$ higher than in tiered-AutoNUMA, 
because the number of sampled pages for profiling in Thermostat is much larger than that in tiered-AutoNUMA. The profiling overhead in Thermostat is 2.5$\times$ higher than in \name, 
because manipulating reserved bits in PTE and counting protection faults in Thermostat is more expensive than scanning PTEs in tiered-AutoNUMA and \name. With tiered-AutoNUMA, the application run time is longer than with \name by 22\%. This indicates that random sampling-based profiling is not as effective as our adaptive profiling. 

\begin{figure}[t!]
    \centering
        \includegraphics[width=\columnwidth]{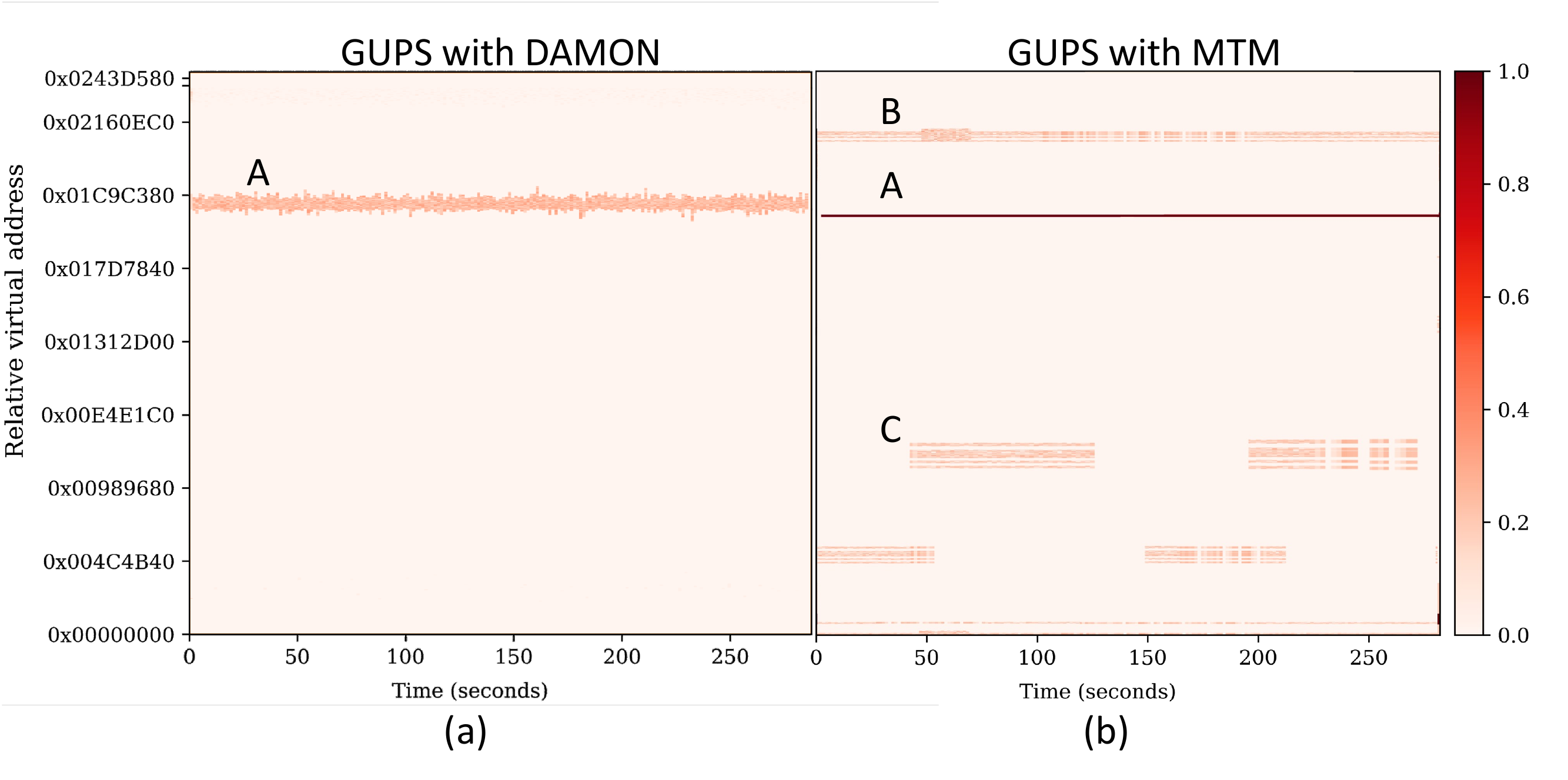}
        \vspace{-20pt}
        \caption{The heatmap of memory accesses in GUPS using (a) DAMON and (b) \name under the same profiling overhead (5\%).}
        \label{fig:heatmap}
\end{figure}


\textbf{Comparison with DAMON.}
We compare DAMON and \name in terms of profiling quality. We use GUPS that randomly accesses 512GB memory with 24 threads. Specifically, 20\% of \textcolor{check}{GUPS'} memory footprint is randomly selected as the hotset. Each thread randomly updates the memory 1M times, and 80\% of them happens in the hotset. 
1M-updates repetitively happens, so that there is variance on hot pages. 
Using knowledge on GUPS, we know there are three hot data objects:  the indexes to access the hotset (labeled as ``A''), the hotset information (labeled as ``B''), and the hotset (labeled as ``C'').


Figure~\ref{fig:heatmap} shows results. (1) \name finds C, while DAMON cannot because of its slow response. (2) \name finds B, but DAMON cannot because its memory regions are initially set based on the virtual memory area tree, which is too coarse-grained to capture B even after splitting regions. (3) DAMON and \name find A, but the scope of A found by \name is correctly narrowed down, which reduces unneeded migration. 

We further evaluate the following profiling techniques and disable them one by one to examine performance difference.

\textbf{Effectiveness of adaptive memory regions.} 
We disable adaptive memory regions but respect the profiling overhead constraint. Figure~\ref{fig:profiling} shows the application execution time is 22\% longer, although the overhead constraint is met. Such a loss indicates that hot memory regions are not effectively identified without adaptive regions and hence placed in slow memory.

\textbf{Effectiveness of adaptive page sampling.} This technique distributes PTE scans between memory regions by using time-consecutive profiling, which includes information on temporal locality. We disable it and randomly distribute PTE scans between memory regions, and observe 21\% performance loss. 

\textbf{Evaluation of profiling overhead control.} We set $\tau_1=\tau_2=0$ (i.e., no merging/splitting memory regions) and observe that the profiling time is increased by 3$\times$ in Figure~\ref{fig:profiling}.


\textbf{Evaluation of performance-counter assistance.} 
MTM-w/o-PEBS in Figure~\ref{fig:profiling} does not use performance counters to guide profiling. It performs worse than \name because PEBS improves profiling quality. For VoltDB, it performs 4\% worse. The overhead of PEBS is less than 1\%.

\subsubsection{Sensitivity Study for Memory Profiling}
\begin{figure}[t!]
    \centering        \includegraphics[width=0.85\columnwidth]{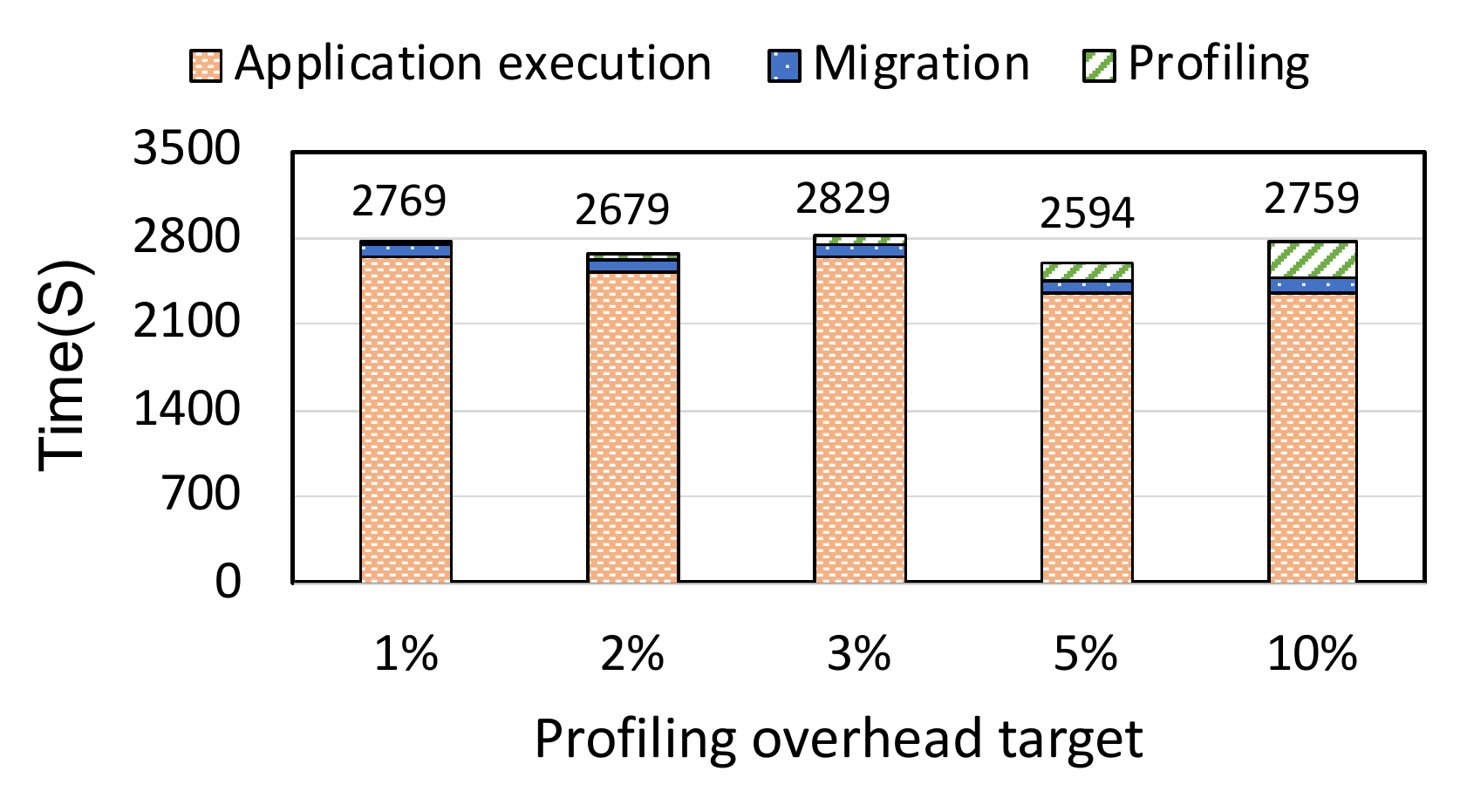}
        \vspace{-5pt} \caption{\textcolor{revision}{Execution time with various profiling overhead targets. We evaluate voltDB with \name.}}
         \vspace{-5pt} \label{fig:overhead}
\end{figure}

\textcolor{revision}{\textbf{Impacts of profiling overhead.} We study the relationship between the profiling overhead and profiling quality. Figure~\ref{fig:overhead} shows the results. We set profiling interval length as 5s, and test a set of profiling overhead targets. As the overhead target increases from 1\% to 10\%, the application execution time is reduced by 12\%. Such performance improvement comes from the improvement of profiling quality when trading profiling overhead (by taking more samples) for profiling quality. However, taking more samples (or using a larger overhead target) does not necessarily lead to better performance. As shown in Figure~\ref{fig:overhead}, the application execution time increases by 7\% as the profiling overhead target increases from 5\% to 10\%. We use 5\% as the profiling overhead target by default, which universally works well for all applications in our study.}



\begin{figure}[t!]
    \centering
        \includegraphics[width=\columnwidth]{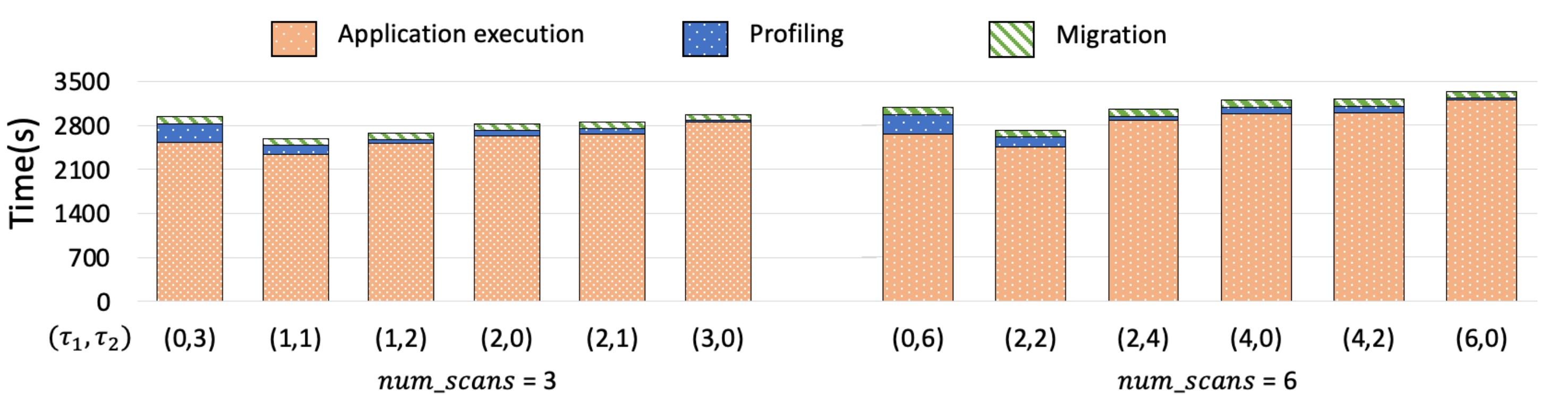}
        \vspace{-10pt}
\caption{\textcolor{revision}{Execution time with various $\tau_1$ and $\tau_2$. We execute voltDB with \name.}}
        \vspace{-5pt}
        \label{fig:t1t2}
\end{figure}

\textcolor{revision}{\textbf{Impacts of profiling thresholds $\tau_1$ and $\tau_2$.}
We study the relationship between memory region merging/split thresholds ($\tau_1$ and $\tau_2$) and profiling quality. The increase of $\tau_1$ leads to aggressive merging of memory regions, and the decrease of  $\tau_2$ leads to aggressive split of memory regions. We change $\tau_1$ and $\tau_2$ and measure performance. 
Figure~\ref{fig:t1t2} shows the results.}


\textcolor{revision}{With $num\_scans$ is set as 3, $\tau_1=1$ and $\tau_2=2$ lead to the best performance, outperforming other configurations of $\tau_1$ and $\tau_2$ by at least 7\%. The execution time of voltDB increases as $\tau_1$ increases. We observe that more aggressive merging of memory region leads to inaccurate profiling results: In the extreme case, when $tau_1=num\_scans$, there is only one memory region. The inaccurate profiling results leads to bad application performance. Both profiling overhead and execution time increase when $\tau_2$ decreases. With aggressive split of memory regions, MTM uses a long time to capture memory access pattern of the application, which increases profiling overhead and loses profiling quality. We observe the same trend when $num\_scans$ is set as 6.}

\begin{figure}[t!]
    \centering
        \includegraphics[width=\columnwidth]{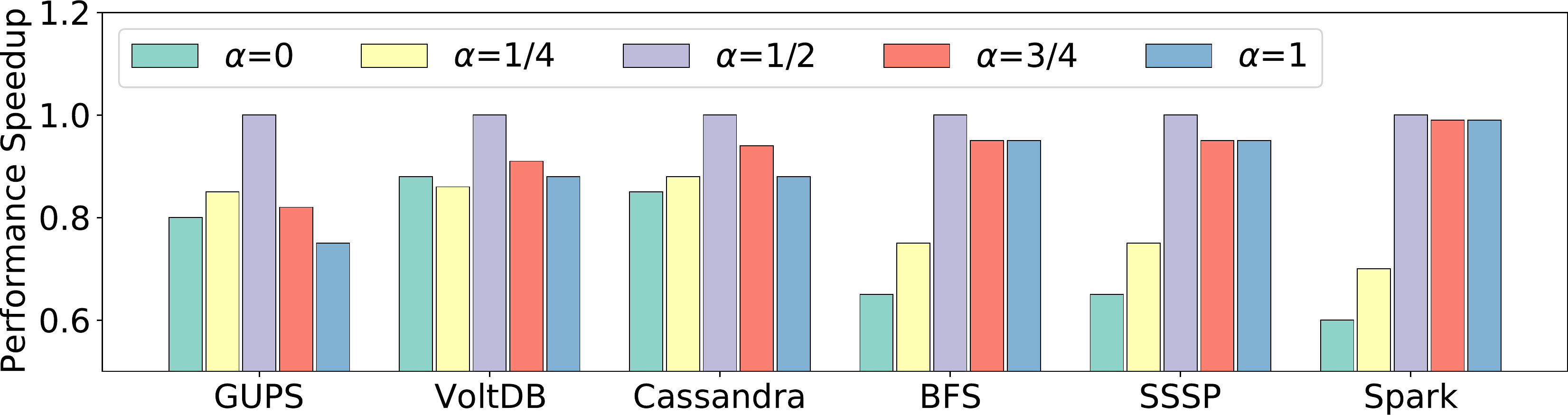}
        \caption{\textcolor{revision}{Performance variance when changing $\alpha$. The performance is normalized by that of the default setting $\alpha=1/2$.}}
\label{fig:alpha}
\end{figure}

\textcolor{revision}{\subsection{Sensitivity study for Migration Strategy}}
\textcolor{revision}{\textbf{Impact of memory promotion threshold $\alpha$.} The memory promotion threshold $\alpha$ is used in Equation~\ref{eq:whi} to balance the contributions of the historical profiling results and current profiling results. When $\alpha=0$, MTM makes migration decisions only based on the historic profiling result. When $\alpha=1$, MTM ignores historic information. We set $\alpha$ with different values for sensitivty study. Figure~\ref{fig:alpha} demonstrates that different applications have different sensitivity to $\alpha$. Using both historic and current profiling results are generally helpful for most of applications (e.g., GUPS, voltDB, Cassandra, BFS and SSSP). In contrast, Spark is not sensitive to the inclusion of historical profiling results. This may be because of lack of temporal locality in Spark.}

\subsection{Migration Mechanism in \name}
\begin{figure}[t!]
    \centering
        \includegraphics[width=\columnwidth]{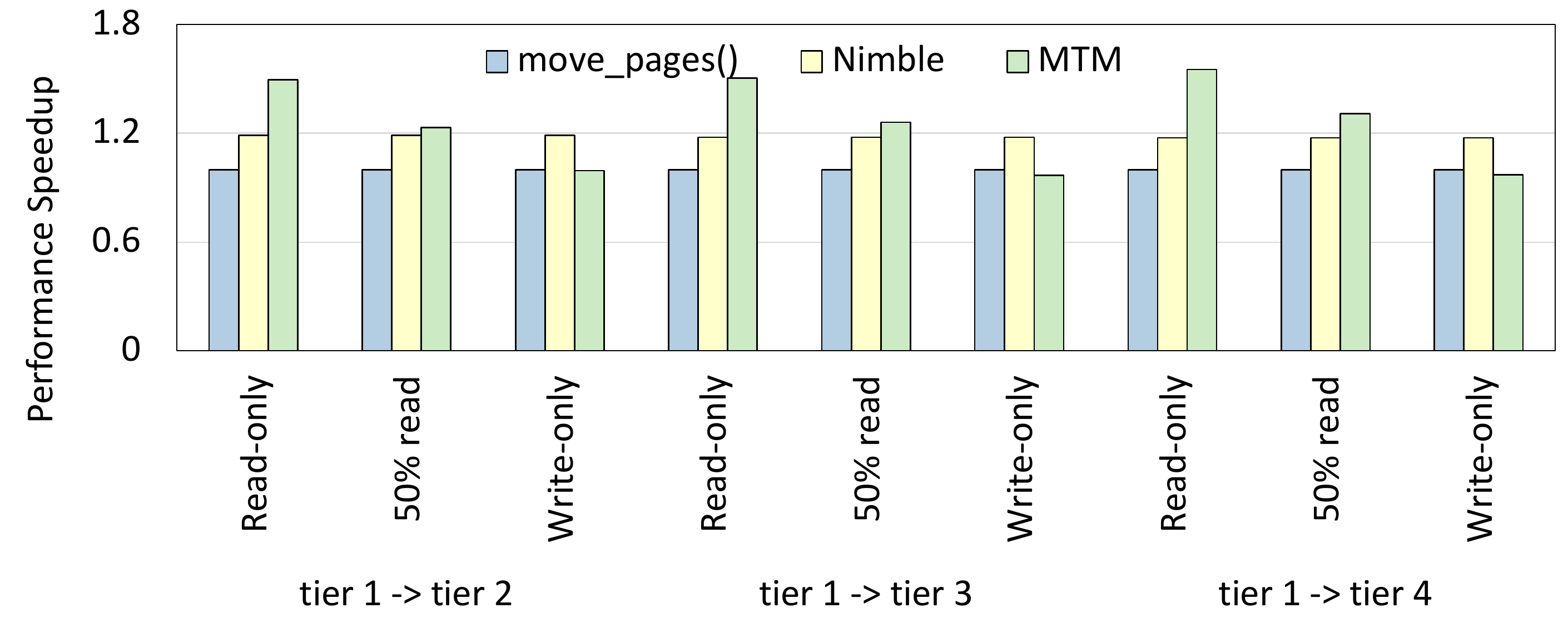}
        \caption{Performance comparison between Nimble, $move\_pages()$ and \name. The performance is normalized by the performance of $move\_pages()$.}
        \label{fig:migration}
\end{figure}

\textbf{Effectiveness of Migration Mechanism}
\label{sec:eval_migration_mechanism}
We use three microbenchmarks to evaluate the migration mechanisms in \name, Nimble~\cite{Yan:2019:NPM:3297858.3304024}, and \textit{move\_pages()} in Linux: sequential read-only, 50\% read (i.e., a sequential read followed by an update on an array element), and 100\% sequential write on a 1GB array. The array is allocated and touched in a memory tier, and then migrated to another. 
Figure~\ref{fig:migration} shows the results. Migrating pages between the tiers 1 and 2, \name's mechanism performs 40\%, 23\%, and -0.5\% better than \textit{move\_pages()}, and performs 26\% 4\% and -6\% better than Nimble, for the three benchmarks respectively. We see the same trend in other tiers. In general, for read-intensive pages, \name's mechanism brings large benefit; 
for write-intensive pages, \name performs similar to \textit{move\_pages()} and Nimble. 

\textcolor{revision}{\textbf{Overhead of user-space page fault handler.} The user-space page fault handler is used in  \textit{move\_memory\_regions()} to track writes and enable asynchronous page copying. We measure its overhead by repeatedly triggering faults. On average, it takes about 40$\mu s$ to handle a page fault. Furthermore, we compare the performance of \textit{move\_memory\_regions()} without the user-space page fault handler (i.e., disable userfd), with it (but handling all userfd on the critical execution path), and with MTM.}

\begin{figure}[t!]
    \centering
        \includegraphics[width=\columnwidth]
        {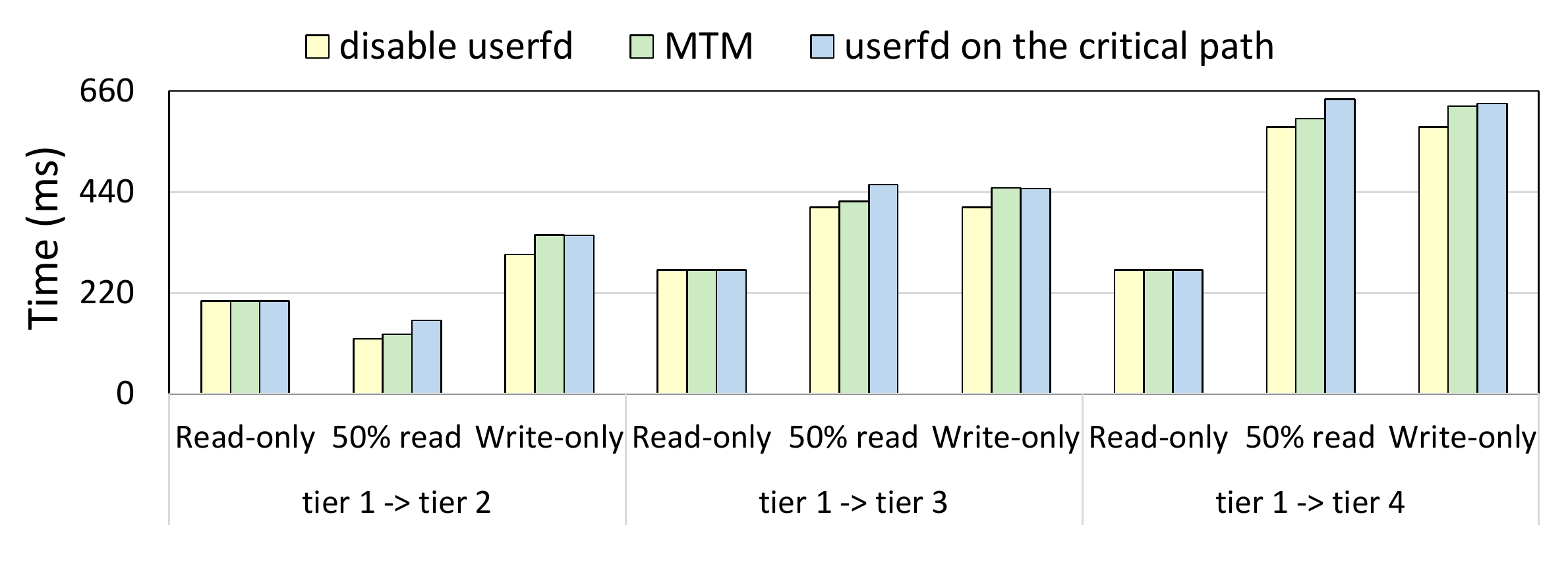}
        \vspace{-15pt}
        \caption{Performance comparison between disabling userfd, MTM and handle userfd on the critical path.}
         \vspace{-5pt}
         \label{fig:overhead_handler}
\end{figure}

\textcolor{revision}{Figure~\ref{fig:overhead_handler} shows the results. We use the same microbenchmarks for evaluation. The read-only microbenchmark does not trigger the user-space page fault handling. In the 50\%-read microbenchmark, a part of the overhead of the user-space page fault handling can be hidden from the critical path. MTM outperforms ``userfd on the critical path'' by 6\% at most. The overhead of the user-space page fault handling in MTM is 3\% on average. In write-only cases, the overhead of the user-space page fault handling are all on the critical path. The overhead of the user-space page fault handling in MTM is 7\% on average.}


\subsection{\name with two-tiered HM}
\begin{figure}[t!]
    \centering
        \includegraphics[width=0.8\columnwidth]{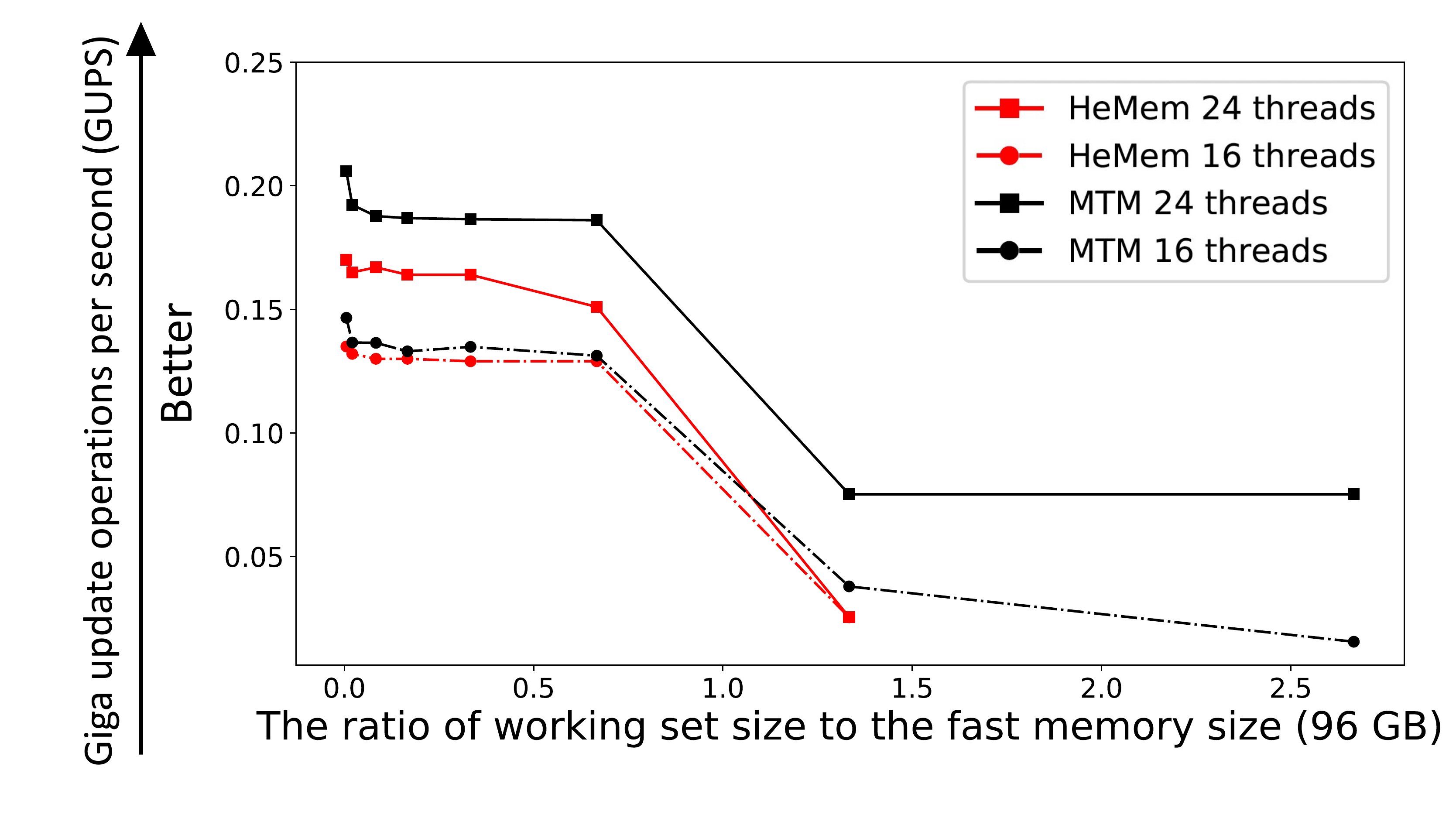}
        \vspace{-10pt}
        \caption{Evaluation of \name on two-tiered HM and comparison with HeMem.}
         \vspace{-5pt}
        \label{fig:hemem}
\end{figure}

The evaluation is performed on a single socket with two tiers, using  GUPS~\cite{gups} as in HeMem~\cite{sosp21_hemem}.
Figure~\ref{fig:hemem} reports using 16 and 24 application threads. The results show that when the working set size fits in the fast memory tier (i.e., the ratio in the \texttt{x} axis is smaller than 1.0), \name performs similarly to HeMem at 16 threads but better at 24 threads. Once the working set size exceeds the fast memory, HeMem fails to sustain performance at 24 threads while \name still sustains higher performance at 24 threads than 16 threads. \name performs better because its profiling method can quickly adapt to changes in memory accesses and identify more hot pages.

\subsection{\textcolor{revision}{\name with Phase-change Workload}}
\textcolor{revision}{we introduce different execution phases into GUPS. Specifically, we concatenate four GUPS execution blocks into a single workload. Different GUPS blocks have different memory access patterns. This allows us to build a phase-change workload with the apriori knowledge on page hotness.}

\begin{figure}[t!]
    \centering
        \includegraphics[width=\columnwidth]{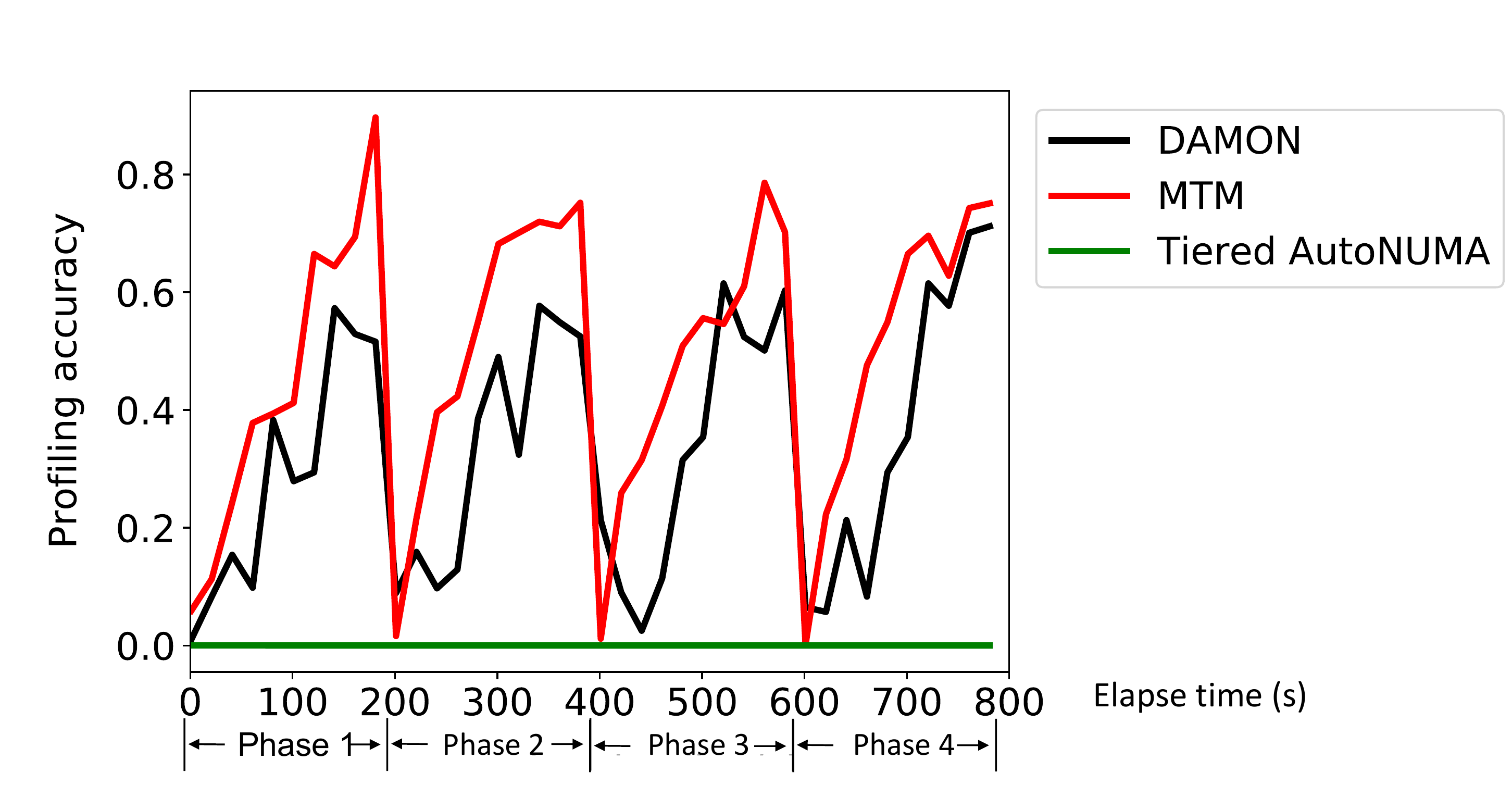}
        
        \caption{\textcolor{revision}{Profiling accuracy with DAMON, \name and Tiered AutoNUMA. We use phase-change GUPS}}
        \label{fig:phase-change}
\end{figure}

\textcolor{revision}{Figure~\ref{fig:phase-change} shows the profiling accuracy of \name, DAMON and Tiered-AutoNUMA. The results indicate that Tiered-AutoNUMA cannot capture phase changes well. The profiling accuracy of \name largely outperforms that of Tiered-AutoNUMA by 2.7$\times$ on average. DAMON can detect phase changes. However, DAMON shows slow response to phase changes. DAMON takes 3X longer time than MTM to achieve 50\% profiling accuracy. This is because DAMON builds memory region randomly. In contrast, with assists from PEBS for adaptive profiling, \name can quickly detect potentially hot memory regions and quickly adjust the memory region size. It achieves 80\% accuracy within 50s in each execution phase.}





\textcolor{revision}{We further compare  execution times for different page migration strategies. 
\name outperforms NUMA frist-touch, HMC, AutoTiereing, and Tiered AutoNUMA by 15\%, 17\%, 23\%,and 26\% respectively.}

\subsection{\textcolor{revision}{\name with Application Co-runs}}
\begin{figure}[t!]
    \centering
        \includegraphics[width=1.05\columnwidth]{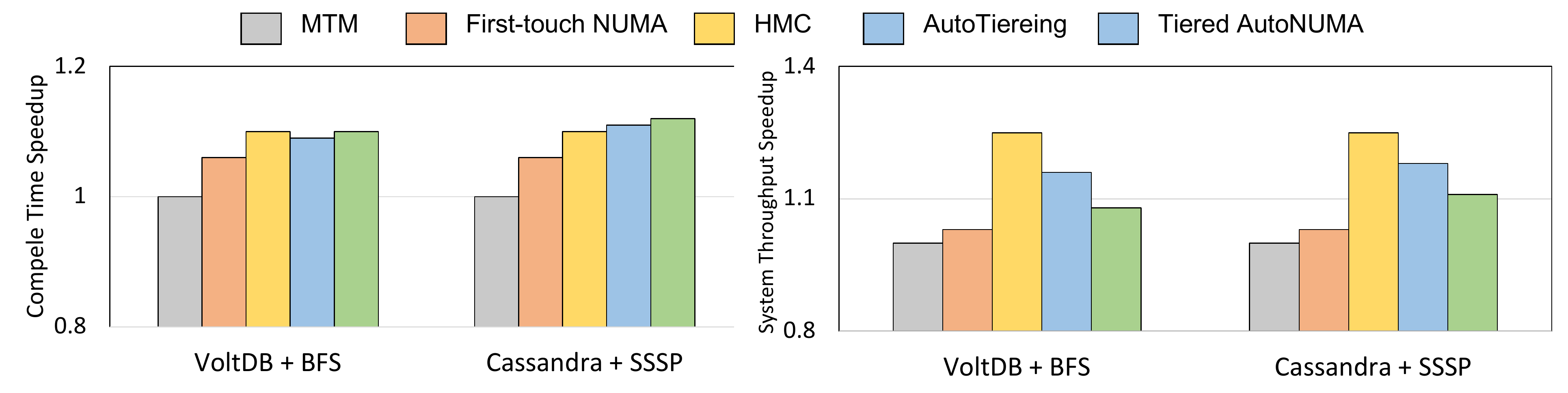}
        \vspace{-5pt}
        \caption{\textcolor{revision}{Completion time and system throughput for application co-run. The results are normalized by those of \name}}
         \vspace{-5pt}
        \label{fig:corun}
\end{figure}

\textcolor{revision}{We launch two applications simultaneously. We report system throughput (i.e., the number of write bytes caused by workload execution per second) and the completion time of co-applications. We test two co-run workloads, i.e., VoltDB with BFS co-run, and Cassandra with SSSP co-run. Figure~\ref{fig:corun} shows the result. MTM outperforms NUMA frist-touch, HMC, AutoTiereing, and Tiered AutoNUMA by 6\%, 12\%, 10\%, 11\% respectively in terms of average completion time. \name  outperforms NUMA frist-touch, HMC, AutoTiereing, and Tiered AutoNUMA by 3\%, 25\%, 18\%, 11\% in system throughput.}



\textcolor{revision}{Currently, \name handles application co-run by profiling each application and making page migration decisions independently, which loses a   global view of the entire system. We leave the page placement optimization across applications as our future work.}

\section{Conclusions}
\vspace{-2pt}
Emerging multi-tiered large memory systems calls for rethinking memory profiling and migration for high performance. We present \name, a page management system customized for large memory systems based on three design principles of improving profiling quality, building a universal page migration policy and huge page awareness. Our extensive evaluation of \name against multiple state-of-the-art solutions shows that \name can largely outperform existing solutions by up to 42\%.

\clearpage

\bibliographystyle{plain}
\bibliography{li, jie}

\begin{thebibliography}{10}

\bibitem{cxl}
{Compute Express Link Industry Members}.
\newblock https://www.computeexpresslink.org/members, 2021.

\bibitem{Agarwal:2017:TAP:3037697.3037706}
Neha Agarwal and Thomas~F. Wenisch.
\newblock {Thermostat: Application-transparent Page Management for Two-tiered
  Main Memory}.
\newblock In {\em International Conference on Architectural Support for
  Programming Languages and Operating Systems}, 2017.

\bibitem{Andrei:2017:SHA:3137765.3137780}
Mihnea Andrei, Christian Lemke, G\"{u}nter Radestock, Robert Schulze, Carsten
  Thiel, Rolando Blanco, Akanksha Meghlan, Muhammad Sharique, Sebastian
  Seifert, Surendra Vishnoi, Daniel Booss, Thomas Peh, Ivan Schreter, Werner
  Thesing, Mehul Wagle, and Thomas Willhalm.
\newblock Sap hana adoption of non-volatile memory.
\newblock {\em Proc. VLDB Endow.}, 10(12), August 2017.

\bibitem{nvmw21:dram-cache}
Julian~T. Angeles, Mark Hildebrand, Venkatesh Akella, and Jason Lowe-Power.
\newblock {Investigating Hardware Caches for Terabyte-scale NVDIMMs}.
\newblock In {\em Annual Non-Volatile Memories Workshop}, 2021.

\bibitem{cassandra}
Apache.
\newblock {Open Source NoSQL Database}.
\newblock \url{https://cassandra.apache.org/}, 2021.

\bibitem{vldb21:ai_pm}
Cheng Chen, Jun Yang, Mian Lu, Taize Wang, Zhao Zheng, Yuqiang Chen, Wenyuan
  Dai, Bingsheng He, Weng-Fai Wong, Guoan Wu, Yuping Zhao, and Andy Rudoff.
\newblock {Optimizing In-Memory Database Engine for AI-Powered Online Decision
  Augmentation Using Persistent Memory}.
\newblock In {\em Proceedings of the VLDB Endowment}, 2021.

\bibitem{ipdps21:dancing}
Jinyoung Choi, Sergey Blagodurov, and Hung-Wei Tseng.
\newblock Dancing in the dark: Profiling for tiered memory.
\newblock In {\em 2021 IEEE International Parallel and Distributed Processing
  Symposium (IPDPS)}, 2021.

\bibitem{ycsb}
Brian~F. Cooper, Adam Silberstein, Erwin Tam, Raghu Ramakrishnan, and Russell
  Sears.
\newblock {Benchmarking Cloud Serving Systems with YCSB}.
\newblock In {\em Proceedings of the 1st ACM Symposium on Cloud Computing},
  2010.

\bibitem{autonuma}
J.~Corbet.
\newblock {{AutoNUMA: the Other Approach to NUMA Scheduling}}.
\newblock http://lwn.net/Articles/488709.

\bibitem{vldb20_sage}
Laxman Dhulipala, Charles McGuffey, Hongbo Kang, Yan Gu, Guy~E. Blelloch,
  Phillip~B. Gibbons, and Julian Shun.
\newblock Sage: Parallel semi-asymmetric graph algorithms for nvrams.
\newblock {\em Proc. VLDB Endow.}, 13(9):1598–1613, May 2020.

\bibitem{kleio:hpdc19}
Thaleia~Dimitra Doudali, Sergey Blagodurov, Abhinav Vishnu, Sudhanva
  Gurumurthi, and Ada Gavrilovska.
\newblock {Kleio: A Hybrid Memory Page Scheduler with Machine Intelligence}.
\newblock In {\em International Symposium on High-Performance Parallel and
  Distributed Computing}, 2019.

\bibitem{ipdps21_cori}
Thaleia~Dimitra Doudali, Daniel Zahka, and Ada Gavrilovska.
\newblock Cori: Dancing to the right beat of periodic data movements over
  hybrid memory systems.
\newblock In {\em 2021 IEEE International Parallel and Distributed Processing
  Symposium (IPDPS)}, pages 350--359, 2021.

\bibitem{optane_utexas19}
Gurbinder Gill, Roshan Dathathri, Loc Hoang, Ramesh Peri, and Keshav Pingali.
\newblock Single machine graph analytics on massive datasets using intel optane
  dc persistent memory.
\newblock {\em Proc. VLDB Endow.}, 13(8):1304–1318, April 2020.

\bibitem{gups}
GUPS.
\newblock {Giga Updates Per Second}.
\newblock \url{http://icl.cs.utk.edu/ projectsfiles/hpcc/RandomAccess/}, 2021.

\bibitem{spark_terasort}
Ewan Higgs.
\newblock {Spark-terasort}.
\newblock \url{https://github.com/ehiggs/spark-terasort}, 2018.

\bibitem{9408179}
Mark Hildebrand, Julian~T. Angeles, Jason Lowe-Power, and Venkatesh Akella.
\newblock {A Case Against Hardware Managed DRAM Caches for NVRAM Based
  Systems}.
\newblock In {\em IEEE International Symposium on Performance Analysis of
  Systems and Software (ISPASS)}, 2021.

\bibitem{AutoTM_asplos20}
Mark Hildebrand, Jawad Khan, Sanjeev Trika, Jason Lowe-Power, and Venkatesh
  Akella.
\newblock {AutoTM: Automatic Tensor Movement in Heterogeneous Memory Systems
  Using Integer Linear Programming}.
\newblock In {\em International Conference on Architectural Support for
  Programming Languages and Operating Systems}, 2020.

\bibitem{Hirofuchi:2016:RHV:2987550.2987570}
Takahiro Hirofuchi and Ryousei Takano.
\newblock Raminate: Hypervisor-based virtualization for hybrid main memory
  systems.
\newblock In {\em Proceedings of the Seventh ACM Symposium on Cloud Computing},
  2016.

\bibitem{amazon_high_mem_inst}
Amazon Inc.
\newblock {Amazon EC2 High Memory Instances with 6, 9, and 12 TB of Memory,
  Perfect for SAP HANA.}
\newblock
  https://aws.amazon.com/blogs/aws/now-available-amazon-ec2-high-memory-instances-with-6-9-and-12-tb-of-memory-perfectfor-sap-hana/,
  September 2018.

\bibitem{intel_mem_optimizer}
Intel.
\newblock {Intel Memory Optimizer}.
\newblock \url{https://github.com/intel/memory-optimizer}, 2019.

\bibitem{mem_optimizer_intel}
Intel.
\newblock {Intel Memory Optimizer}.
\newblock \url{https://github.com/intel/memory-optimizer}, 2019.

\bibitem{tiered-autonuma}
Intel.
\newblock {Autonuma: Optimize Memory Placement in Memory Tiering System}.
\newblock \url{https://lwn.net/Articles/803663/}, 2020.

\bibitem{memory_tiering}
Intel.
\newblock {Intel Memory Tiering}.
\newblock \url{https://lwn.net/Articles/802544/}, 2021.

\bibitem{intel_pcm}
Intel.
\newblock {Intel Processor Counter Monitor}.
\newblock \url{https://github.com/opcm/pcm}, 2021.

\bibitem{Kannan:2017:HOD:3079856.3080245}
Sudarsun Kannan, Ada Gavrilovska, Vishal Gupta, and Karsten Schwan.
\newblock {HeteroOS: OS Design for Heterogeneous Memory Management in
  Datacenter}.
\newblock In {\em International Symposium on Computer Architecture}, 2017.

\bibitem{asplos21:kloc}
Sudarsun Kannan, Yujie Ren, and Abhishek Bhattacharjee.
\newblock {KLOCs: Kernel-Level Object Contexts for HeterogeneousMemory
  Systems}.
\newblock In {\em International Conference on Architectural Support for
  Programming Languages and Operating Systems}, 2021.

\bibitem{atc21_autotiering}
Jonghyeon Kim, Wonkyo Choe, and Jeongseob Ahn.
\newblock {Exploring the Design Space of Page Management for Multi-Tiered
  Memory Systems}.
\newblock In {\em USENIX Annual Technical Conference}, 2021.

\bibitem{atc21:autotiering}
Jonghyeon Kim, Wonkyo Choe, and Jeongseob Ahn.
\newblock {Exploring the Design Space of Page Management for Multi-Tiered
  Memory Systems}.
\newblock In {\em 2021 {USENIX} Annual Technical Conference ({USENIX} {ATC}
  21)}, 2021.

\bibitem{asplos19_softwarefarmem}
Andres Lagar-Cavilla, Junwhan Ahn, Suleiman Souhlal, Neha Agarwal, Radoslaw
  Burny, Shakeel Butt, Jichuan Chang, Ashwin Chaugule, Nan Deng, Junaid Shahid,
  Greg Thelen, Kamil~Adam Yurtsever, Yu~Zhao, and Parthasarathy Ranganathan.
\newblock Software-defined far memory in warehouse-scale computers.
\newblock In {\em Proceedings of the Twenty-Fourth International Conference on
  Architectural Support for Programming Languages and Operating Systems}, 2019.

\bibitem{tpcc}
Scott~T. Leutenegger and Daniel Dias.
\newblock {A Modeling Study of the TPC-C Benchmark}.
\newblock In {\em SIGMOD Record}, 1993.

\bibitem{ms_cxl_2022}
Huaicheng Li, Daniel~S. Berger, Stanko Novakovic, Lisa Hsu, Dan Ernst, Pantea
  Zardoshti, Monish Shah, Ishwar Agarwal, Mark~D. Hill, Marcus Fontoura, and
  Ricardo Bianchini.
\newblock {First-generation Memory Disaggregation for Cloud Platforms}, 2022.

\bibitem{autonuma_balance}
Linux.
\newblock {Automatic NUMA Balancing}.
\newblock
  \url{https://www.linux-kvm.org/images/7/75/01x07b-NumaAutobalancing.pdf},
  2014.

\bibitem{tpds19_tieredmem}
Lei Liu, Shengjie Yang, Lu~Peng, and Xinyu Li.
\newblock Hierarchical hybrid memory management in os for tiered memory
  systems.
\newblock {\em IEEE Transactions on Parallel and Distributed Systems},
  30(10):2223--2236, 2019.

\bibitem{ATC20_leap}
Hasan~Al Maruf and Mosharaf Chowdhury.
\newblock Effectively prefetching remote memory with leap.
\newblock In {\em 2020 {USENIX} Annual Technical Conference ({USENIX} {ATC}
  20)}, pages 843--857. {USENIX} Association, July 2020.

\bibitem{meta_tpp}
Hasan~Al Maruf, Hao Wang, Abhishek Dhanotia, Johannes Weiner, Niket Agarwal,
  Pallab Bhattacharya, Chris Petersen, Mosharaf Chowdhury, Shobhit Kanaujia,
  and Prakash Chauhan.
\newblock {TPP: Transparent Page Placement for CXL-Enabled Tiered Memory},
  2022.

\bibitem{pm-octree:sc17}
Bao Nguyen, Hua Tan, and Xuechen Zhang.
\newblock Large-scale adaptive mesh simulations through non-volatile
  byte-addressable memory.
\newblock In {\em Proceedings of the International Conference for High
  Performance Computing, Networking, Storage and Analysis}, SC ’17, 2017.

\bibitem{hpdc22:daos}
SeongJae Park, Madhuparna Bhowmik, and Alexandru Uta.
\newblock {DAOS: Data Access-Aware Operating System}.
\newblock In {\em Proceedings of International Symposium on High-Performance
  Parallel and Distributed Computing (HPDC)}, 2022.

\bibitem{middleware19_profiling}
SeongJae Park, Yunjae Lee, and Heon~Y. Yeom.
\newblock Profiling dynamic data access patterns with controlled overhead and
  quality.
\newblock In {\em Proceedings of the 20th International Middleware Conference
  Industrial Track}, Middleware '19, page 1–7, New York, NY, USA, 2019.
  Association for Computing Machinery.

\bibitem{peng2018siena}
I.~B. {Peng} and J.~S. {Vetter}.
\newblock Siena: Exploring the design space of heterogeneous memory systems.
\newblock In {\em SC18: International Conference for High Performance
  Computing, Networking, Storage and Analysis}, 2018.

\bibitem{peng2018graphphi}
Zhen Peng, Alexander Powell, Bo~Wu, Tekin Bicer, and Bin Ren.
\newblock Graphphi: efficient parallel graph processing on emerging
  throughput-oriented architectures.
\newblock In {\em Proceedings of the 27th International Conference on Parallel
  Architectures and Compilation Techniques}, pages 1--14, 2018.

\bibitem{sosp21_hemem}
Amanda Raybuck, Tim Stamler, Wei Zhang, Mattan Erez, and Simon Peter.
\newblock {HeMem: Scalable Tiered Memory Management for Big Data Applications
  and Real NVM}.
\newblock In {\em Proceedings of the ACM SIGOPS 28th Symposium on Operating
  Systems Principles}, 2021.

\bibitem{hpca21_sentinel}
Jie Ren, Jiaolin Luo, Kai Wu, Minjia Zhang, Hyeran Jeon, and Dong Li.
\newblock {Sentinel: Efficient Tensor Migration and Allocation on Heterogeneous
  Memory Systems for Deep Learning}.
\newblock In {\em International Symposium on High Performance Computer
  Architecture (HPCA)}, 2021.

\bibitem{neurips20:hm-ann}
Jie Ren, Minjia Zhang, and Dong Li.
\newblock {HM-ANN: Efficient Billion-Point Nearest Neighbor Search on
  Heterogeneous Memory}.
\newblock In {\em Conference on Neural Information Processing Systems
  (NeurIPS)}, 2020.

\bibitem{damon}
{{SeongJae Park}}.
\newblock {{DAMON: Data Access Monitor}}.
\newblock https://sjp38.github.io/post/damon/.

\bibitem{Optane:blogreview}
Billy Tallis.
\newblock {The Intel Optane Memory (SSD) Preview: 32GB of Kaby Lake Caching}.
\newblock
  \url{http://www.anandtech.com/show/11210/the-intel-optane-memory-ssd-review-32gb-of-kaby-lake-caching}.

\bibitem{thp}
{{The Linux Kernel User’s and Administrator’s Guide}}.
\newblock {{Transparent Hugepage Support}}.
\newblock https://www.kernel.org/doc/html/latest/admin-guide/mm/transhuge.html.

\bibitem{voltdb}
voltDB.
\newblock {voltDB}.
\newblock \url{https://www.voltdb.com/}, 2021.

\bibitem{pldi19:panthera}
Chenxi Wang, Huimin Cui, Ting Cao, John Zigman, Haris Volos, Onur Mutlu, Fang
  Lv, Xiaobing Feng, and Guoqing~Harry Xu.
\newblock Panthera: Holistic memory management for big data processing over
  hybrid memories.
\newblock In {\em Proceedings of the 40th ACM SIGPLAN Conference on Programming
  Language Design and Implementation}, PLDI 2019, 2019.

\bibitem{ics21:memoization}
Zhen Xie, Wenqian Dong, Jie Liu, Ivy Peng, Yanbao Ma, and Dong Li.
\newblock {MD-HM: Memoization-based Molecular Dynamics Simulations on Big
  Memory System}.
\newblock In {\em {International Conference on Supercomputing (ICS)}}, 2021.

\bibitem{Yan:2019:NPM:3297858.3304024}
Zi~Yan, Daniel Lustig, David Nellans, and Abhishek Bhattacharjee.
\newblock {Nimble Page Management for Tiered Memory Systems}.
\newblock In {\em International Conference on Architectural Support for
  Programming Languages and Operating Systems}, 2019.

\bibitem{ucsd_otpane:fast2020}
Jian Yang, Juno Kim, Morteza Hoseinzadeh, Joseph Izraelevitz, and Steve
  Swanson.
\newblock An empirical guide to the behavior and use of scalable persistent
  memory.
\newblock In {\em 18th {USENIX} Conference on File and Storage Technologies
  ({FAST} 20)}, 2020.

\bibitem{spark}
Matei Zaharia, Mosharaf Chowdhury, Tathagata Das, Ankur Dave, Justin Ma, Murphy
  McCauly, Michael~J Franklin, Scott Shenker, and Ion Stoica.
\newblock Resilient distributed datasets: A fault-tolerant abstraction for
  in-memory cluster computing.
\newblock In {\em 9th $\{$USENIX$\}$ Symposium on Networked Systems Design and
  Implementation ($\{$NSDI$\}$ 12)}, pages 15--28, 2012.

\end{thebibliography}

\end{document}